\documentstyle[aps,graphicx,epsf,rotate]{revtex}

\begin{document}
\draft
\title{A mean field stochastic theory for species-rich assembled communities}
\author{Alan McKane$^{1}$, David Alonso$^{2,3}$ and Ricard V. Sol\'e$^{2,4}$}
\address{$^{1}$Department of Theoretical Physics, University of Manchester,  
Manchester M13 9PL, UK \\ 
$^{2}$Complex Systems Research Group, Departament of Physics, 
FEN, \\
Universitat Polit\`ecnica de Catalunya, Campus Nord B4, 
08034 Barcelona, Spain \\
$^{3}$Department of Ecology, Facultat de Biologia, Universitat de Barcelona,
Diagonal 645, 08045 Barcelona, Spain \\
$^{4}$Santa Fe Institute, 1399 Hyde Park Road, New Mexico 87501, USA}
\maketitle      
\begin{abstract}
A dynamical model of an ecological community is analyzed within a ``mean-field
approximation" in which one of the species interacts with the combination of
all of the other species in the community. Within this approximation the
model may be formulated as a master equation describing a one-step stochastic 
process. The stationary distribution is obtained in closed form and is shown 
to reduce to a logseries or lognormal distribution, depending on the values
that the parameters describing the model take on. A hyperbolic relationship
between the connectance of the matrix of interspecies interactions and the
average number
of species, exists for a range of parameter values. The time evolution of 
the model at short and intermediate times is analyzed using van Kampen's 
approximation, which is valid when the number of individuals in the community
is large. Good agreement with numerical simulations is found. The large time
behavior, and the approach to the stationary state, is obtained by solving
the equation for the generating function of the probability distribution.
The analytical results which follow from the analysis are also in good 
agreement with direct simulations of the model. 
\end{abstract}

\section{Introduction}

The steady accumulation of data on all aspects of the very diverse ecosystems 
that exist on Earth has revealed a number of generic features~\cite{eco}. 
Examples include (i) in species-rich ecosystems, the 
number of species, $S(n)$, with $n$ individuals following a power-law, 
$S(n) \approx n^{-\gamma}$, where $\gamma$ is close to one~\cite{eco,i}, 
(ii) a relation between the number of species in the 
ecosystem, $S$, and the connectance, $C^{*}$ --- defined as the number of 
predator-prey links between pairs of species divided by the total possible 
number of links --- which has the hyperbolic form $C^{*} \approx 
S^{-1 + \eta}$, with $\eta \in [0,1]$~\cite{ii}, and (iii) other 
power-law distributions concerning the extinction of species, for instance, 
where the lifetime of species, $T$, appears to be well-described by the 
distribution $N(T) \approx T^{-\theta}$, with $\theta$ between 1.1 and 
1.6~\cite{iii}. There is an urgent need for models of ecosystems to be 
developed which will allow the underlying mechanisms which lead to these 
regularities to be understood. These models need to be defined for an arbitrary
number of species, have a set of rules specifying the interaction between 
pairs of species which is reasonably simple and based on general features such 
as the competition between species, and have a stochastic element to reflect 
the randomness of events which are inherent in real systems. In Ref \cite{I} 
a model of this type was introduced in order to investigate the generic 
features outlined above. An analysis of the model was begun in that paper, 
where both numerical and analytical work showed predictions of the model to be 
in agreement with field data. Here we present a more detailed analysis of the 
model, using a variety of techniques, and compare the results of this analysis 
with that from real ecosystems. We begin by defining the model.

The ecosystem under study is taken to have $N$ individuals and $S$ possible 
species. It is modeled as a directed graph with the nodes labelled by 
$i =1,...,S$ representing the species, and the links representing the
(predator-prey) interaction between the species at the two nodes being joined.
This interaction is assumed to be given by a single real number, denoted by
$\Omega_{ij}$ for the link to $j$ from $i$. Thus, the interaction between the
species in the ecosystem is completely specified by the $S \times S$ real
matrix $\bf \Omega$. Links from a node to itself are not allowed and
therefore this matrix has zero entries on the diagonal. The antisymmetric
matrix $S_{ij} \equiv \Omega_{ij} - \Omega_{ji}$ has a more direct 
interpretation as the ``score" of species $i$ against species $j$:
\begin{itemize}
\item If $S_{ij} > 0$, then $j$ acts as a resource for $i$. 

\item If $S_{ij} = 0$, there is no interaction between $i$ and $j$.

\item If $S_{ij} < 0$, then $i$ acts as a resource for $j$. 
\end{itemize}

Modelling multispecies ecosystems involving species-species interactions or
connections of this type, has a long history~\cite{odum}. Originally, 
population dynamics equations, such as the Lotka-Volterra equations, were 
written down for two species and then for many species. If one imagines 
studying the equations near to any fixed point that might exist, it is 
permissable to linearize about the fixed points, and the entire model is then 
specified by a single $S \times S$ matrix --- the stability matrix. Whereas, 
for systems involving two species it might be useful to calculate this matrix 
in terms of the original parameters of the model, for systems of many species 
there are simply too many parameters and so the emphasis changed to trying to 
investigate general properties that such matrices might have. An obvious, but 
crude, assumption that the entries were random, was first investigated by 
May~\cite{may1} who found that the connectivity of the matrix was important in 
determining its stability properties. Since then, this has remained a central 
issue in ecology~\cite{sta}, as has the study of the abstract theory of species
connected by a complex network of interactions~\cite{jai}.

Having described the basic idea we will use in our approach, we now need
to specify the interaction matrix $\bf \Omega$. Since, the connectivity 
seems to emerge as an important quantity in both theoretical and experimental
studies we will assign a fixed conductivity $C$ to ${\bf \Omega}$, so that 
we may study how properties of the system change as $C$ is varied. Other than
this, and the fact that the diagonal entries are zero, we will not impose any
other restrictions on $\bf \Omega$. We now have to define the dynamics. The
goal is to define a set of rules which is simple, but which builds up a complex
model ecosystem, after a sufficiently long time, showing the non-trivial
emergent behavior mentioned at the beginning of this section. We do this by
assigning the (off-diagonal) entries of $\bf \Omega$, in a purely random way at 
$t = 0$, and updating the system at discrete time steps as follows. At each
time step:
\begin{itemize}
\item[1.]With probability $(1 - \mu )$, pick two individuals at random. Suppose 
they belong to species $i$ and $j$ and that $S_{ij} \neq 0$. Replace the 
individual belonging to the species which has a negative score against the 
other species by a new individual of the more successful species. So, for 
example, if $S_{ij} > 0$, the total number of individuals belonging to species 
$i$ goes up by one, and the total belonging to species $j$ goes down by one. If
$S_{ij} = 0$, no action is taken.
\item[2.]With probability $\mu$, pick an individual at random. Replace it by
another individual of {\em any} of the $S$ species.
\end{itemize}
These rules have an obvious interpretation. The first simply ensures that the
most successful species, in the sense of the ones having the highest scores, 
grow at the expense of the less successful ones. However, if the dynamics 
consisted only of this rule, then eventually all species but one would go
extinct. Therefore, a second rule has to be introduced in order to
obtain a diverse ecosystem. The simplest choice is to violate the first rule
occasionally, by giving even unsuccessful species an opportunity through purely
random events. This is best not thought of as a mutation or speciation, but as
an immigration event from an area outside the ecosystem under study. 

We have not specified the initial distribution of the entries in $\bf \Omega$
and there is a certain amount of freedom regarding this choice. In our
simulations we have chosen the $\Omega_{ij}\, (i \neq j)$ randomly 
from a uniform distribution on $[0,1]$, but any other choice is equally valid,
since only the sign of $S_{ij}$ is important. Since the probability that 
$\Omega_{ij} = \Omega_{ji} \neq 0$ is vanishingly small, it is almost certain 
that if the condition $S_{ij} = 0$ mentioned in rule 1 holds, then both 
$\Omega_{ij}$ and $\Omega_{ji}$ are zero, and species $i$ and $j$ are not
connected by a predator-prey relationship. Since the probability of any matrix
element being zero is $1 - C$, the probability that both $\Omega_{ij}$ and 
$\Omega_{ji}$ are zero is $(1 - C)^{2}$, and from what has been said
above, the probability that $S_{ij}$ is non-zero is 
$C^{*} \equiv 1 - (1 - C)^{2}$. 

For those simulations that start with no species in the system, a generalized 
$(S+1)\times(S+1)$ real matrix $\bf \Omega '$ with an extra row and column 
denoted by 0 needs to be introduced. Then, if $S_{i0}>0$, empty space acts as a 
resource for $i$. If $S_{i0}<0$, then species $i$ fails to invade empty space. 
So, on average, the expected number of species that can actually invade empty
space is $SC/2$ and the number of species in the pool that can never interact
with empty space is $(1-C)S$. 

What has been described above is a strongly interacting, stochastic 
multispecies model and as such is 
extremely difficult to study analytically. It is, however, relatively easy to
simulate, given the straightforward nature of the algorithm described above, and
we will discuss the results of extensive numerical studies in later sections of
this paper. It would still be very useful to have some approximate treatment
available which would, at the very least, help to suggest forms which could be
use to fit the data and simulations. Fortunately, we can do much better than
this. A mean field theory of the above model yields a master equation which can
be analysed using a number of standard techniques. Much of the paper will be
concerned with the derivation of these results and their subsequent
interpretation. 

The plan of the paper is as follows. In section II we derive the master
equation within the mean field approximation and in section III we investigate 
the nature of the stationary state. The time-dependent properties are the 
subject of the next two sections: within a Gaussian approximation in section IV
and a more general study in section V. We conclude with a summary of the work
presented in the paper in section VI. There are two appendices: Appendix A and 
Appendix B contain technical details which are used to derive some of the 
results in sections III and V respectively.

\section{Master equation}

In this section we will derive a master equation which approximately describes 
the complex dynamics introduced in section I. The key simplification is the use
of a type of mean field theory. We focus on one of the $S$ species, which we
shall call species $A$. The other $(S - 1)$ species are no longer distinguished
as separate species and are simply lumped together and denoted as species $B$.  
The $B$ species will be regarded as some kind of average species --- a kind
of effective background population --- with which species $A$ interacts. There
are various assumptions inherent in this approach. For instance, that the rate
of reproduction is the same for all species, so that a typical species ($A$)
can be picked out as representative. It does however, reduce the model to one
in which just two species are interacting, namely $A$ and non-$A = B$. It is
now relatively straightforward to derive a master equation which describes the
dynamics of this process.

To derive this equation, first suppose that $\mu = 0$ for simplicity. Then only
rule 1 is in operation. In picking two individuals from a set of $N$
individuals of the $S$ species, the following situations arise: (a) both 
individuals belong to species $A$, (b) one belongs to $A$ and the other to 
$B$ and (c) both individuals belong to species $B$. In cases (a) and (c) there
is no action taken under rule 1. The probability of case (b) occurring is the
sum of the probability that first an $A$ is selected and then a $B$ and the
probability that a $B$ is selected and then an $A$:
\begin{displaymath}
\frac{n}{N}\,\left( 1 - \frac{n - 1}{N - 1} \right) \ + \
\left( 1 - \frac{n}{N} \right)\,\left( \frac{n}{N - 1} \right) \ = 
\frac{2n}{N}\,\left( \frac{N - n}{N - 1} \right) \, , 
\end{displaymath}
where $n$ is the number of individuals of species $A$ in the ecosystem. We 
now have to focus on the quantity $S_{AB}$ in order to implement the rule. 
The probability that it is non-zero is $C^{*}$ and we would expect that, on
average, half of the events the individual from species $A$ will have a higher 
score than the individual from species $B$ i.e. $S_{AB} < 0$ and the other 
half of the events to have $S_{AB} < 0$. Therefore, the probability that at 
each time step the number of species $A$ increases by one is
\begin{displaymath}
W(n + 1 | n ) = C^{*}\frac{n}{N}\,\left( \frac{N - n}{N - 1} \right) \, , 
\end{displaymath}
and the probability that at each time step the number of species $A$ decreases 
by one is
\begin{displaymath}
W(n - 1 | n ) = C^{*}\frac{n}{N}\,\left( \frac{N - n}{N - 1} \right) \, . 
\end{displaymath}
We will show shortly that this process leads to a stationary probability 
distribution which is non-zero only if $n = 0$ or $n = N$, that is, only if
either species $A$ dies out completely or dominates completely. The second rule
ensures that some diversity is retained. 

So suppose that we now include the second rule. The transition probabilities
above have now to be multiplied by $(1 - \mu)$ and those involving the second
rule will involve a factor $\mu$. Specifically, the probability that the 
individual picked, when implementing the second rule, is replaced by an 
individual of a {\em given} species is $\mu/S$. If we ask that this given 
species is $A$, then since the probability that the individual picked belongs 
to species $B$ is $(1 - n/N)$, the additional probability due to rule 2 that
at each time step the number of species $A$ is increased by one is
\begin{displaymath}
\frac{\mu}{S}\,\left( 1 - \frac{n}{N} \right)\, .
\end{displaymath}
Similarly, the probability that an individual is replaced by an individual of a
different species is $\mu(S - 1)/S$. Since the probability that the individual 
picked belongs to species $A$ is $n/N$, the additional probability that at each
time step the number of species $A$ is decreased by one is 
\begin{displaymath}
\frac{\mu}{S}\,(S - 1)\,\frac{n}{N}\, .
\end{displaymath}
Putting the two rules together, gives the one-step transition probabilities 
$g_n \equiv W(n+1\vert n)$ and $r_n \equiv W(n-1\vert n)$ as
\begin{equation}
g_n = C^* (1-\mu){n \over N} \left( {N-n \over N-1} \right) 
+ {\mu \over S} (1-{n\over N})\ ,
\label{gn}
\end{equation}
and
\begin{equation}
r_n = C^* (1-\mu){n \over N} \left( {N-n \over N-1} \right) 
+ {\mu \over S} (S-1){n\over N}\ . 
\label{rn}
\end{equation}
We can now write down a master equation describing this one-step stochastic
process~\cite{van,gar}. If $P(n,t)$ is the probability of species $A$ having $n$
individuals at time $t$, the master equation takes the form
\begin{eqnarray}
{dP(n,t) \over  dt} & = & r_{n+1} P(n+1,t) + g_{n-1} P(n-1,t) \nonumber \\
& - & (r_n + g_n) P(n,t)\,.
\label{master}
\end{eqnarray}
Equation (\ref{master}) is only valid for values of $n$ not on the boundary
(i.e. for $n \neq 0$ and $n \neq N$); for these values special equations have
to be written reflecting the fact that no transitions out of the region $[0,N]$ 
are allowed. However, from (\ref{gn}) and (\ref{rn}) we see that $g_N=0$ and 
$r_0=0$, and if additionally we define $r_{N+1}=0$ and $g_{-1}=0$, then 
(\ref{master}) holds for all $n = 0,1,...,N$. To completely specify the system
we also need to give an initial condition, which will typically have the form
$P(n,0) = \delta_{n,m}$ for some non-negative integer $m$. 

We will end this section by determining the stationary probability
distribution, $P_s(n)$. Setting $dP(n)/dt=0$, one obtains
\begin{equation}
r_{n+1} P_s(n+1) - g_n P_s(n) = r_n P_s(n) - g_{n-1} P_s(n-1)\, . 
\label{current}
\end{equation}
This is true for all $n$, which implies that $r_n P_s(n) - g_{n-1} P_s(n-1) =
J$, where $J$ is a constant. Applying the boundary condition at $n = 0$, we 
find that $J = 0$ and therefore
\begin{equation}
r_n P_s(n) = g_{n-1} P_s(n-1)\ ; \ n = 0,1,...,N\, . 
\label{balance}
\end{equation}
If $\mu \ne 0$, then $r_n \neq 0$ for all $n$ such that $0 < n \leq N$, and
therefore
\begin{equation}
P_s(n) = \frac{g_{n-1}\, g_{n-2} ...\, g_0}{r_n\, r_{n-1} ...\, r_1} P_s(0)\ ; 
\ n = 1,...,N\, .
\label{Ps}
\end{equation}
The constant $P_s(0)$ can be determined from the normalization condition
\begin{equation}
\sum_{n=0}^N P_s(n) = P_s(0) + \sum_{n>0} P_s(n) = 1\, :
\label{normalization}
\end{equation}
\begin{equation}
(P_s(0))^{-1} = 1 + \sum_{n=1}^{N} \frac{g_{n-1}\, g_{n-2} ...\, g_0}{r_n\, 
r_{n-1} ...\, r_1}\, .  
\label{invPs(0)}
\end{equation}
At this point it is convenient to introduce a set of combinations of the
constants of the model which will appear frequently in the analysis. These are:
\begin{displaymath}
\mu^*=\mu/[(1-\mu)SC^{*}]\ \ , \ \ \lambda^*=\mu^*(N-1)
\end{displaymath}
\begin{equation}
{\rm and} \ \ \nu^*=N+\mu^*(N-1)(S-1)\, .
\label{newconsts}
\end{equation}
The transition probabilities (\ref{gn}) and (\ref{rn}) may now be written in
the more compact form
\begin{eqnarray}
g_n & = & \frac{C^{*}(1 - \mu)}{N(N - 1 )}\, (N - n)(\lambda^{*} + n)\ \ 
{\rm and} \nonumber \\ \nonumber \\
r_n & = & \frac{C^{*}(1 - \mu)}{N(N - 1 )}\, n(\nu^{*} - n)\, .
\label{gnrn}
\end{eqnarray}
Substituting (\ref{gnrn}) into (\ref{invPs(0)}) gives
\begin{eqnarray}
(P_s(0))^{-1} & = & \sum_{n=0}^N {N \choose n} 
{\Gamma (n + \lambda^*) \over \Gamma(\lambda^*)} {\Gamma (\nu^* - n)
\over \Gamma(\nu^*)} \nonumber \\ \nonumber \\
& = & \sum_{n=0}^N {N \choose n} (-1)^n 
{\Gamma (n + \lambda^*) \over \Gamma(\lambda^*)} {\Gamma (1-\nu^*)
\over \Gamma(n + 1 - \nu^*)}\ .
\label{sum}
\end{eqnarray}
This sum takes the form of a Jacobi polynomial 
$P_N^{(\alpha, \beta)}(x)$~\cite{abr}, with $\alpha =-\nu^*$,  
$\beta =\lambda^*+\nu^*-(N+1)$ and $x = -1$, which can itself be expressed in 
terms of gamma functions for this value of $x$. So using (\ref{Ps}) we find
\begin{equation}
P_s(n) = {N \choose n} {\Gamma (n + \lambda^*) \over \Gamma(\lambda^*)} 
{\Gamma (\nu^* - n) \over \Gamma(\nu^* - N)} {\Gamma (\lambda^* + \nu^* - N) 
\over \Gamma(\lambda^* + \nu^*)}\ .
\label{Ps(n)_gamma}
\end{equation}
In various intermediate expressions we have assumed that $\nu^{*}$ is not an 
integer, but this final result is well defined for all meaningful ranges of 
the parameters, since from (\ref{newconsts}) we can see that $\nu^{*} > N$
and $\lambda^{*} > 0$. Moreover $P_s(n) > 0$ for all $n = 0,1,...,N$ and 
$\sum_n P_s(n) = 1$ by construction. By introducing the beta function
$\beta(p,q)=\Gamma(p) \Gamma(q) / \Gamma(p+q)$, the stationary, normalized 
solution can be written in the more compact form 
\begin{equation}
P_s(n) = {N \choose n} {\beta(n+\lambda^*, \nu^*-n) \over \beta(\lambda^*,
\nu^*-N) }\ .  
\label{Ps(n)_beta}
\end{equation}
Finally, if $\mu = 0$, $r_N = 0$, so (\ref{Ps}) no longer holds for $n = N$. 
Using the result that $g_0 = 0$ in this case, one finds that $P_s(n) = 0$
for $n = 1,...,N-1$. By normalization we can write $P_s(0) = {\cal C}$ and 
$P_s(N) = 1-{\cal C}$, where $\cal C$ is a constant. So, as mentioned earlier, 
either species $A$ is the only surviving species or it goes extinct. In other
words, in the stationary state only one species survives. Since all species are
assumed identical, it follows that when $\mu = 0$
\begin{equation}
P_s(0) = \frac{1}{S}\ \ ; \ P_s(N) = 1 - \frac{1}{S}\ \ ; \ P_s(n) = 0\ 
{\rm for}\ 0 < n < N\, .
\label{muiszero}
\end{equation}
Although we have obtained the exact solution for the stationary distribution 
(\ref{Ps(n)_gamma}) in terms of nothing more complicated than gamma functions,
we still need to simplify it if we are to compare the result with data. In the
next section, we will derive simpler forms for the stationary probability 
distribution which are valid in different regions of the parameter space of 
the model, and compare these with simulations. 

\section{stationary state}

We have so far been discussing the stationary state from the point of view of
a time-independent solution of the master equation. But let us now ask the 
question in a biological context: are ecological communities in stable 
equilibria? Although is obvious that environmental variability and chance have 
a great impact on ecosystems, some well-defined, time-independent, patterns 
arise when natural ecosystems are observed. The model we have introduced 
reaches a well-established dynamic stationary state which allows us to study 
some of these patterns. A particular example of interest is the way that 
individuals are distributed among species. In any island where colonization
from the mainland and local extinction take place, a dynamical equilibrium 
between these two processes is reached~\cite{MacArthurWilson67}. In these 
situations our model applies and can help to understand the patterns observed.

Highly diverse ecological communities are formed when a large number of 
different species are present. The estimation and characterization of such 
biological diversity is not only a central issue in theoretical ecology, but 
also a question of practical concern for nature reserve design and conservation
biology in general. In any ecological community species vary considerably in
the number of individuals that belong to that species. Some species are very
difficult to find because they are very rare. Some of them are extremely common.
How are individuals distributed among species? What factors affect this 
distribution? The classic way of studying this topic is by means of species 
abundance relations --- the ``relations between abundance and the number of 
species possessing that abundance"~\cite{May75}. Different types of species 
abundance relations have been used to fit to real species abundance data. Some 
of them have been justified on theoretical grounds (see~\cite{eco} for a 
review). One of the most widely used species abundance distribution was first 
discussed by Fisher, Corbet and Williams in 1943~\cite{Fisher43}. The 
distribution is defined by two parameters $x$ and $\alpha$:
\begin{equation}
S(n)=\frac{\alpha x^{n}}{n}\,,
\label{log-series_0}
\end{equation}
where $S(n)$ is the number of species having $n$ individuals. Since  
(\ref{log-series_0}) summed over $n$ gives a logarithm, this is known as the
logseries distribution. It is very common as a sampling distribution in the 
ecological literature, although it has also been derived on theoretical 
grounds~\cite{Engen96,Pueyo00}.

The abundance distribution that has received more attention from ecologists,
however, was introduced by Preston in one of the most influential papers on 
ecological theory~\cite{Preston48,Preston62}. As May remarks ``theory and 
observation points to its ubiquity once $S\gg 1$, when relative abundances 
must be governed by the conjunction of a variety of independent 
factors"~\cite{May75}. The distribution is the lognormal distribution, so
called because the logarithm of species abundances is normally distributed:
\begin{equation}
S(R)=S(n_{0})\exp(-\frac{R^{2}}{2\rho ^{2}})\,,
\label{log-normal_0}
\end{equation}
where, following Preston's definitions, $R=\log_{2}(n/n_{0})$ is a logarithmic
measure of the abundance in relation to $n_{0}$ --- the abundance value where
the distribution has its maximum. So, $S(R)dR$ is the number of species having 
their logarithmic relative abundance between $R$ and $R+dR$. Notice that both 
equations (\ref{log-series_0}) and (\ref{log-normal_0}) must be divided by the 
total number of species to be properly understood as estimations of the
probability distribution function.

In this section we want to compare the exact result for $P_{s}(n)$ with
these two distributions --- the most widely used abundance distributions in the
ecological arena. We will derive simpler forms for the stationary probability 
distribution which are valid in different regions of the parameter space of 
the model, and compare these with simulations of the original model (that is, 
without making the mean-field approximation). Lognormal and logseries 
distributions will naturally emerge for different well-defined immigration 
regimes. In a forthcoming paper we will analyse a large quantity of species 
abundance data from different ecological communities in detail. 

We begin by discussing one situation in which the logseries distribution
occurs. It turns out to be convenient to rewrite the result (\ref{Ps(n)_gamma})
for $P_s(n)$ by breaking it down into three separate parts:
\begin{equation}
P_s(n) = {\cal F}_{1}(n)\, {\cal F}_{2}(N)\, {\cal F}_{3}(n,N)\, ,
\label{effs}
\end{equation}
where
\begin{eqnarray}
{\cal F}_{1}(n) & = & \frac{\Gamma(n + \lambda^{*})}{n!\,\Gamma(\lambda^{*})},
\nonumber \\ \nonumber \\
{\cal F}_{2}(N) & = & \frac{\Gamma(\nu^{*})}{\Gamma(\lambda^{*} + \nu^{*})}
\frac{\Gamma([\nu^{*} - N] + \lambda^{*})}{\Gamma(\nu^{*} - N)} \ \ 
\left( = P_s(0) \right), \nonumber \\ \nonumber \\
{\rm and}\ \ {\cal F}_{3}(n,N) & = & \frac{N!}{(N - n)!} 
\frac{\Gamma(\nu^{*} - n)}
{\Gamma(\nu^{*})}\,. 
\label{defeffs}
\end{eqnarray}
We will now give a simpler form for each of these expressions, being careful 
to state the range of validity of our approximations in each case. Details of
the derivation of these results is given in Appendix A. The non-trivial
behavior occurs for relatively small values of $n$, so in what follows we
will only be interested in values of $n$ up to $n_{\rm max}$, where
$n_{\rm max} \ll N$ and $N \gg 1$. We will also suppose that there are 
many possible species: $S \gg 1$.

From Eqns. (\ref{simple_f1}) and (\ref{simple_f3}) we have that
\begin{equation}
{\cal F}_{1}(n) \approx \frac{\lambda^{*}}{n} \ \ ; \ n \agt 1 \ ; \ 
\lambda^{*} \ll \frac{1}{\ln n_{\rm max}}
\label{f1}
\end{equation}
and
\begin{equation}
{\cal F}_{3}(n,N) \approx \exp{ \left( -n\lambda^{*}S/N \right)}\ \ ; 
\ \lambda^{*}S \ll N\ \ ; \ \lambda^{*}S \ll \left(\frac{N}
{n_{\rm max}}\right)^{2}\, .
\label{f3}
\end{equation}
Therefore (\ref{effs})-(\ref{f3}) give
\begin{equation}
P_s(n) = {\cal K} n^{-1} \exp ( - n \mu^* S)\ , 
\label{logseries}
\end{equation}
where ${\cal K} \equiv \lambda^{*}P_s(0)$. This is the logseries 
(\ref{log-series_0}) with $x = e^{-\mu^{*}S}$ and expressed as the fraction of
species represented by $n$ individuals in the steady state. Note that 
(\ref{log-series_0}), by contrast, gives the absolute number of species with a
given abundance $n$.
 
\begin{figure}
{\par\centering \resizebox*{1\columnwidth}{!}{\rotatebox{270}
{\includegraphics{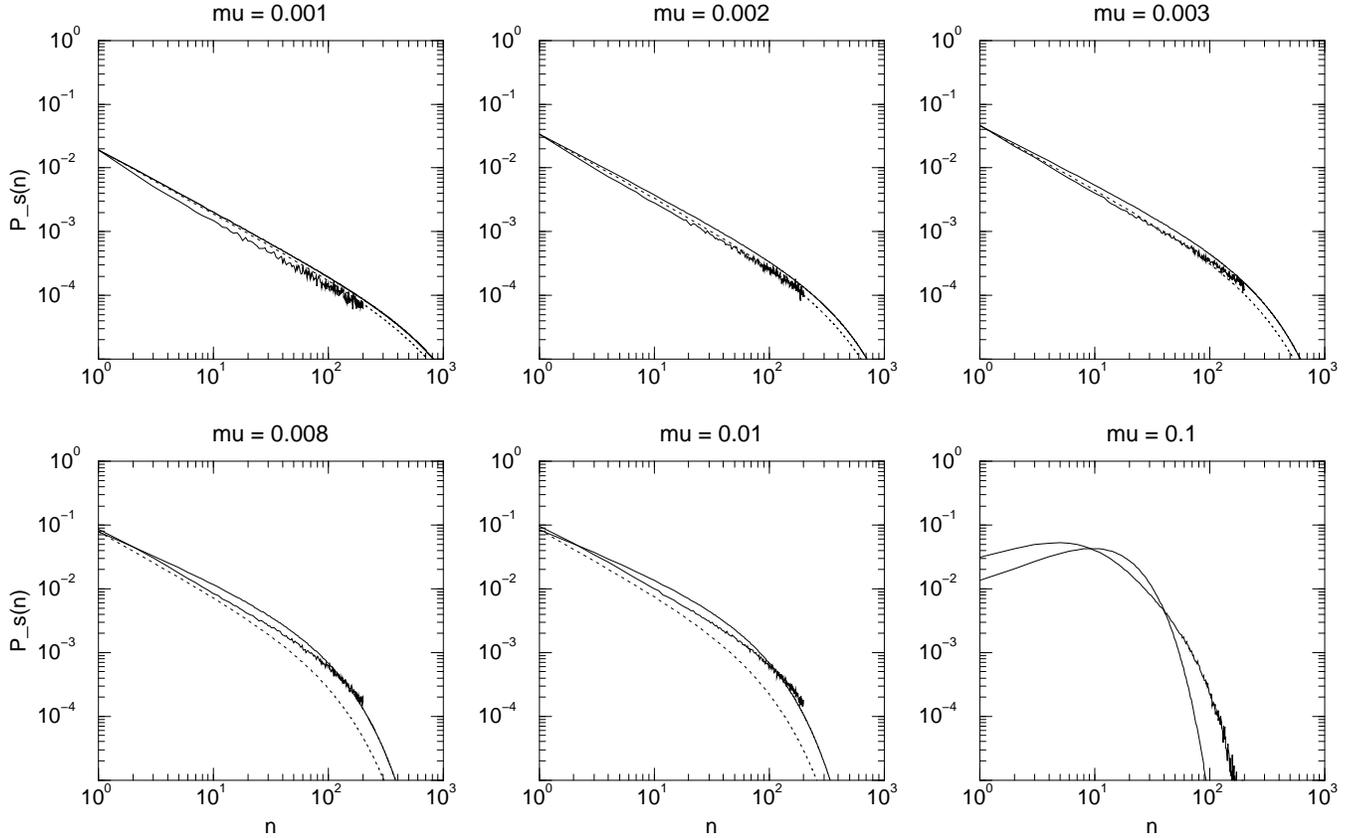}}} \par}
\caption{\label{IBM1} Stationary probability distribution $P_{s}(n)$ obtained 
from the exact solution (\ref{Ps(n)_beta})\ (solid line) and the logseries 
approximation (\ref{logseries})\ (dotted line), with the simplest form for the 
multiplicative factor $\cal{K}$ (\ref{calK}), for different immigration values.
In each plot the species relative abundance distribution resulting from the 
individual based model is also shown (noisy solid line).}
\end{figure}

Since $S < N$, the condition 
$\lambda^{*}S \ll N$ is redundant when the stronger condition 
$\lambda^{*} \ll 1/\ln n_{\rm max}$ is imposed. Therefore, (\ref{logseries})
holds when
\begin{equation}
1 \alt n \le n_{\rm max} \ll \frac{N}{\sqrt{\lambda^{*}S}}\ \ \ {\rm and} \ \ \
\lambda^{*} \ll \frac{1}{\ln n_{\rm max}}\, .
\label{conditions}
\end{equation}
To find an approximate form for
${\cal K}$, we use (\ref{simple_f2}) which gives an approximate form for 
$P_s(0) = {\cal F}_{2}(N)$. Under the very reasonable conditions that 
$\lambda^{*}$ is much less than $N/S$, $\sqrt{N}$ and $S$, but with 
$\lambda^{*}S \agt 1$, we find 
\begin{equation}
{\cal K} \approx \lambda^{*}\, (\mu^{*}S)^{\lambda^{*}}\, .
\label{calK}
\end{equation}
Fig \ref{IBM1} shows the results of different simulations which have been 
performed for increasing values of the immigration parameter. In order to 
calculate the species relative abundance distribution an ensemble average has 
been performed. For each plot a collection of 2000 replicas has been simulated.
For each replica the probability distribution $P(n,t)$ has been calculated 
after 500000 simulation time steps. In Fig \ref{IBM1}, the $\lambda^{*}$ values
increase from $0.033$ when $\mu =0.001$ to $3.3$ when $\mu =0.1$. The last
three plots do not show such a good match with the logseries approximation
as the first three. Even in the upper three plots, where there appears to be
a good fit with the logseries, we would only expect a complete match 
for $1 < n \alt 10$. From (\ref{conditions}), we should bear in mind that this 
is only expected to be true as long as $\lambda^{*}\ll 1/\ln n_{\rm max}$. 
For instance, $\lambda^{*} = 0.22$ for $\mu =10^{-2}$ 
(for the parameter set $N=5000$, $S=300$ and $C=0.5$) 
and $1/\ln n_{\rm max}$ is $0.434$ when $n_{\rm max}=10$.

\begin{figure}
\centering
{\par\centering \resizebox*{1\columnwidth}{!}{\rotatebox{270}
{\includegraphics{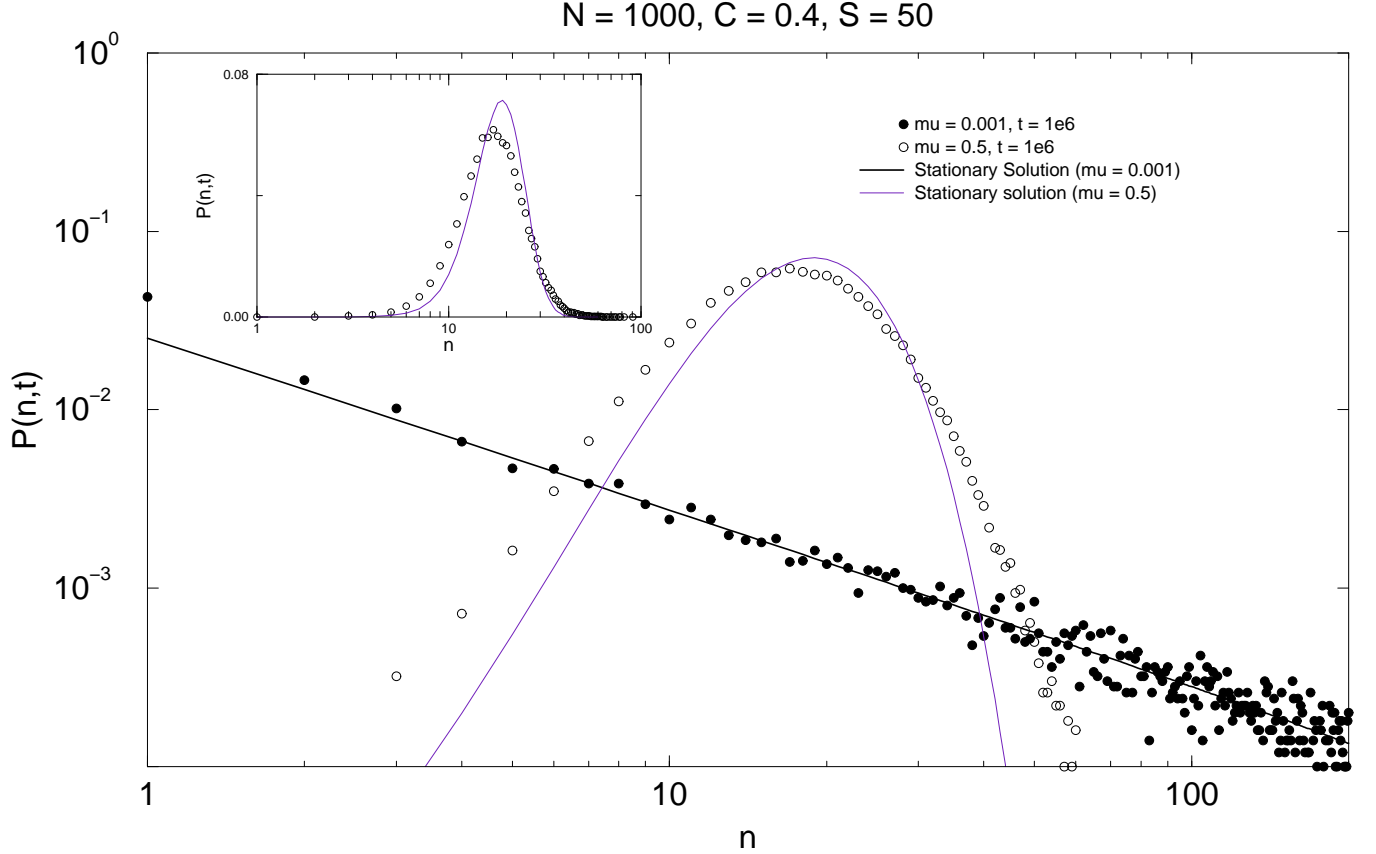}}} \par}
\caption{\label{IBM} Abundance distributions for two individual-based 
simulations are presented for two extreme values of the immigration parameter. 
Time is measured in simulation time units. The stationary distribution 
$P_{s}(n)$ is also shown in both cases.}
\end{figure}

In Fig \ref{IBM} two simulation results are displayed. The stationary solution 
is also shown in both cases for comparison purposes. The stationary solution is
calculated numerically either by direct application of equation 
(\ref{Ps(n)_gamma}), as has been done in Fig \ref{IBM1}, or by means of an
algorithm that can find the stationary  probability distribution of any 
one-step stochastic process if it exists, as in Fig \ref{IBM}. This algorithm
is based on the subroutine TRIDAG~\cite{Press88}. To describe it, 
we first write the master equation (\ref{master}) in the more general form: 
\begin{equation}
{dP(n,t) \over  dt} = \sum_{n \neq n'}\,{\cal W}_{n,n'}\,P(n',t) 
- \sum_{n \neq n'}\,{\cal W}_{n',n}\,P(n,t)\,,
\label{master_gen}
\end{equation}
where $r_{n}={\cal W}_{n-1,n}$ and $g_{n}={\cal W}_{n+1,n}$. If we now 
introduce 
$W_{n,n'} = (1 - \delta_{n,n'})\,{\cal W}_{n,n'} - \delta_{n,n'}\,
\sum_{n'' \neq n}{\cal W}_{n'',n}$ and the vector 
$\vec{P}(t) = (P(1,t),\ldots,P(N,t))$, (\ref{master_gen}) may be written in 
the matrix form
\begin{equation}
\frac{d}{dt}{\vec P}= W \cdot {\vec P}\,.
\label{matrix}
\end{equation}
Finding the stationary stationary distribution $P_{s}(n)$ --- the vector 
${\vec P}_{s}=(P_{s}(0),\ldots ,P_{s}(N))$ --- is then equivalent to solving a 
system of $N+1$ linear equations $W \cdot {\vec P}_{s} = 0$ in $N+1$ unknowns:
$P_{s}(n),\; n=0,\ldots,N$. In any one-step stochastic process the matrix $W$
is tridiagonal. Our algorithm takes advantage of this feature to solve the
system.

\begin{figure}
{\par\centering \resizebox*{10cm}{8cm}{\includegraphics{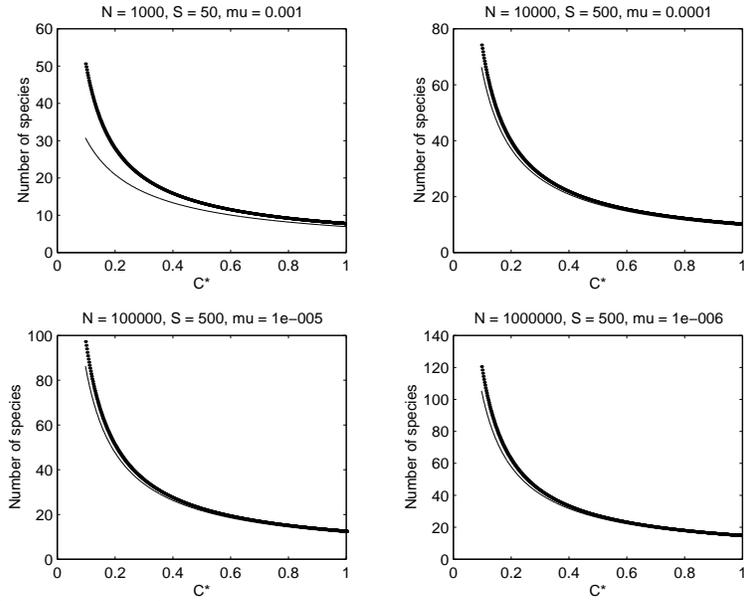}} \par}
\caption{\label{Conn_0} Species-connectivity relationship for different 
parameter values. The expected number of species in the system in the steady 
state is shown plotted against the connectance $C^{*}$. Notice that as long 
as $\lambda ^{*}S$ remains close to but slightly greater than 1, a very good 
agreement is shown between the exact form (\ref{defavs})\ (single line in the 
plot) and the hyperbolic species-connectivity relationship (\ref{CS})\ (double 
line).}
\end{figure}

Another quantity which is useful in comparing model predictions to data is the
average number of species in the stationary state, which we denote by 
$\langle S \rangle$. Let $P_{i}(n)$ be the probability that there are $n$ 
individuals of species $i$ in the ecosystem. Therefore the probability that
there is at least one individual of species $i$ is $1 - P_{i}(0)$ and so
the average number of species is 
$\langle S \rangle = \sum_{i} (1 - P_{i}(0))$. Within the mean field 
approximation $P_{i}(0)$ is the same for all species $i$: $P_{i}(0) = P_s(0)$
(the subscript $s$ denotes ``stationary", as before), so that 
\begin{equation}
\langle S \rangle = \sum^{S}_{i=1} (1 - P_{s}(0)) = (1 - P_{s}(0))S\, .
\label{defavs}
\end{equation}
Under the conditions $\lambda^{*}S \agt 1$, $\lambda^{*} \ll \epsilon$ and
$|\ln C^{*}| \ll |\ln \mu|$, we show in Appendix A that (see eq.
(\ref{simple_S}))
\begin{equation}
\langle S \rangle \sim \left( C^{*} \right)^{-1 + \epsilon}\, ,
\label{CS}
\end{equation}
where $\epsilon^{-1} \approx |\ln \mu |$ (see Fig \ref{Conn_0}). Inverting 
this relationship gives $C^{*} \sim \langle S \rangle^{-1+\eta}$ with 
$\eta = \epsilon/(\epsilon - 1)$. The condition $\lambda^{*}S \agt 1$ is
essentially equivalent to $\mu N \agt 1$ (for a connectance that is not
too small). For systems of interest $N$ is very large, and hence $\mu$
must typically be very small if the hyperbolic relation (\ref{CS}) is
to hold. Such a tiny value of $\mu$ means that $\epsilon$ and hence
$\eta$, will be close to zero. The form of the relationship between 
$C^{*}$ and $\langle S \rangle$ when the immigration parameter has a 
larger value will be discussed elsewhere.

\begin{figure}
{\par\centering \resizebox*{1\columnwidth}{!}{\rotatebox{270}
{\includegraphics{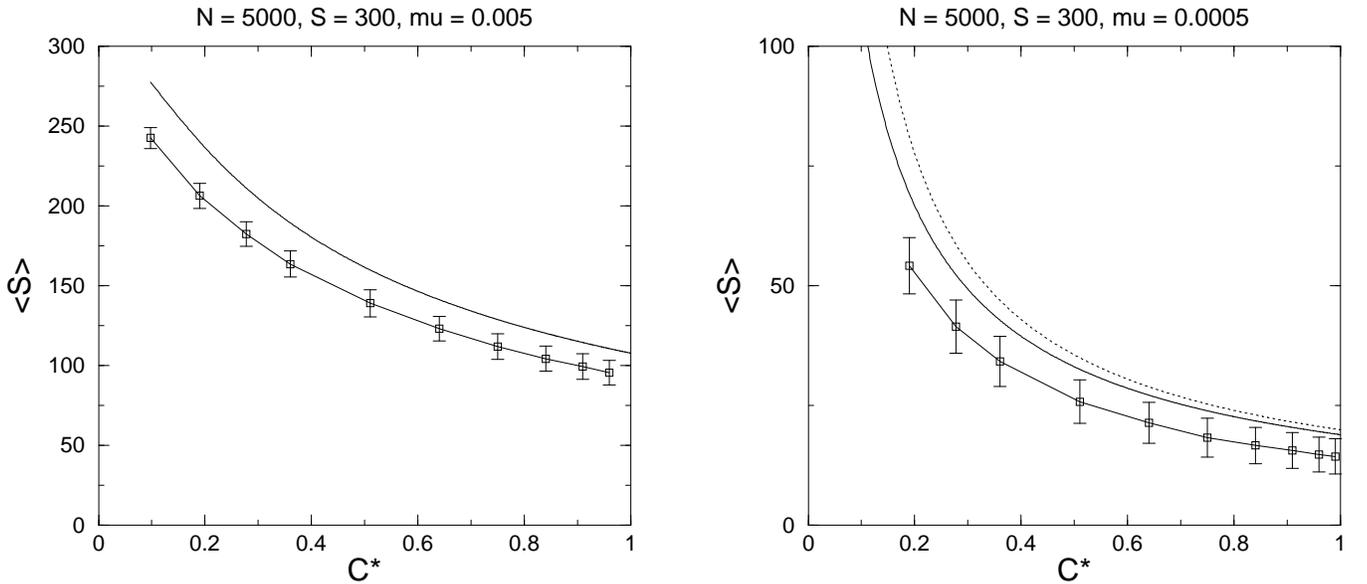}}} \par}
\caption{\label{Conn} Species-connectivity relationship. Two simulations are 
presented. Standard deviations from the ensemble average value are shown. 
In the plot on the left, $\lambda ^{*}S\simeq 25$ --- too high a value to fit 
the approximation given by (\ref{CS}). So in this case the only plotted curve 
is the exact mean field relation (\ref{defavs}). In the plot on the right, 
conditions needed to apply (\ref{CS}) are quite well fulfilled and two curves 
are plotted: the exact mean field relation (solid line) and the approximation 
(dotted line).}
\end{figure}

In Fig \ref{Conn} the species-connectivity relationship calculated from the 
individual modelling approach is shown. After carrying out 600000 simulation
steps, a 1000 ensemble average was calculated for each connectivity value. The
initial condition is the empty system. Although our mean field approximation 
captures the essentials of the hyperbolic-like behavior of the 
species-connectivity relationship, there is a systematic deviation from the 
mean field prediction in the simulated curves. 

\begin{figure}
{\par\centering \resizebox*{10cm}{8cm}{\includegraphics{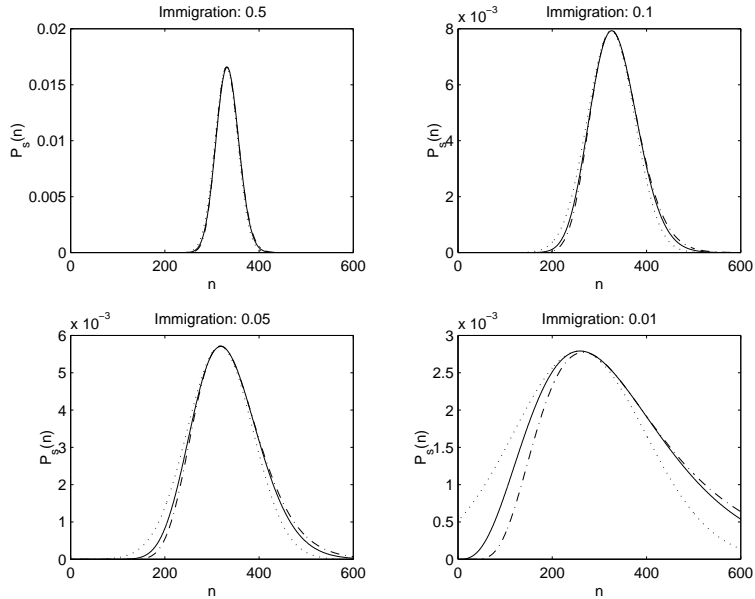}} \par}
\caption{\label{fig_3} Lognormal approximation (the long-dashed curves) for 
the exact stationary solution (\ref{Ps(n)_beta})\ (solid line curves). The 
parameter values that have been used are $N=100000$, $S=300$ and $C=0.5$. 
The Gaussian approximation (dotted curves) is also shown.}
\end{figure}

To sum up, the exact solution given in (\ref{Ps(n)_beta}) admits a logseries
representation for low immigration regimes. For these low immigration values
a hyperbolic-like relation is also observed between the mean number of species
in the stationary state and the connectivity level given by the {\it trophic} 
relationships pre-defined in the community matrix, $\bf \Omega$. We will now
argue that the exact stationary distribution probability is also very well 
approximated by a lognormal distribution for intermediate to high immigration 
regimes as shown in Figs \ref{fig_3} and \ref{IBM}. 

The idea behind the analysis we will present, is to find at which values of 
$n$, if any, $P_{s}(n)$ has a maximum. We then expand $P_{s}(n)$ about this 
maximum to see to what extent this function can be analytically described by a 
lognormal distribution.  

First of all, from Fig \ref{fig_3}, we can see that $P_{s}(n)$ may admit a 
Gaussian representation for some parameter values. So let us look at this case
first, before discussing the lognormal. To investigate for what parameter values
this may occur, we first find the position of the maximum of $P_s(n)$. It is
more convenient to consider $\ln P_{s}(n)$ rather than $P_{s}(n)$. From 
(\ref{Ps(n)_gamma}), we get:
\begin{equation}
\frac{d\ln P_{s}(n)}{dn}=-\psi (n+1)+\psi (n+\lambda ^{*})-\psi 
(\nu ^{*}-n)+\psi (N-n+1)\,,
\label{d_gamma}
\end{equation}
where
\begin{displaymath}
\psi (z) \equiv \frac{d\ln \Gamma (z)}{dz}\,.
\end{displaymath}
Setting (\ref{d_gamma}) equal to zero gives the maximum value of $P_{s}(n)$
at $n=\hat{n}$. If all arguments of the psi-functions can be considered
to be large enough, which is true if $n\ll N$ but reasonably large (e.g.
$n \agt 100$), these functions can be approximated using 
$\psi(z) \sim \ln z$~\cite{abr}. So,
\begin{equation}
\frac{d\ln P_{s}(n)}{dn}\approx -\ln (n+1)+\ln (n+\lambda ^{*})-
\ln (\nu ^{*}-n)+\ln (N-n+1)=0,
\label{d_gamma_1}
\end{equation}
from which one finds that the maximum is given by:
\begin{equation}
\hat{n}=\frac{(N+2)(\lambda ^{*}-1)}{\lambda ^{*}S-2}-1\,.
\label{maximum}
\end{equation}
From (\ref{maximum}) we can see that if $\lambda ^{*}S<2$ (very low immigration
regimes), the numerator and denominator are both negative and a maximum exists.
However it is inadmissible, since it violates the condition $\hat{n} \ll N$.
Therefore, a necessary condition for the existence of a maximum, $\hat{n}$,
is that $\lambda ^{*}>1$. 

Now, we perform a Taylor expansion of (\ref{Ps(n)_gamma}) about $\hat{n}$
to quadratic order. If $n=\hat{n}+\delta n$, then for $\delta n$ small:
\begin{displaymath}
\ln P_{s}(n)=\ln P_{s}(\hat{n})+\frac{1}{2}\left. \frac{d^{2}\ln P_{s}(n)}
{dn^{2}}\right| _{n=\hat{n}}(n-\hat{n})^{2}+O(\delta n)^{3}\,.
\end{displaymath}
Since $\hat{n}$ is a maximum, $d^{2}\ln P_{s}(n)/dn^{2}|_{n=\hat{n}}<0$, and
so we set this equal to $-1/\sigma ^{2}$. Then ignoring the $O(\delta n)^{3}$
terms and exponentiating gives
\begin{equation}
P_{s}(n)=P_{s}(\hat{n})\exp (-\frac{(n-\hat{n})^{2}}{2\sigma ^{2}})\,.
\label{normal}
\end{equation}
Under this approximation ($n\ll N$, but reasonably large) it is not very
difficult to derive an analytical expression for the variance: 
\begin{equation}
\sigma =\frac{\sqrt{\hat{n}^{2}+(\lambda ^{*}+1)\hat{n}+\lambda ^{*}}}
{\sqrt{\lambda ^{*}-1}}\,.
\label{normal_variance}
\end{equation}

Since the Gaussian distribution is completely specified by its first two
cumulants, fixing $\hat{n}$ and $\sigma$ given by (\ref{maximum}) and 
(\ref{normal_variance}) respectively, determines the entire curve. The 
dotted lines in Fig \ref{fig_3} show this curve i.e. (\ref{normal}) with
the two parameters fixed by (\ref{maximum}) and (\ref{normal_variance}).
The upper two plots show very good agreement with the exact mean field
approximation; in the lower two plots the agreement is not so good. 

To approximate the exact stationary solution as a lognormal distribution 
(\ref{log-normal_0}), we will proceed in a similar way. Equation 
(\ref{log-normal_0}) can also be written dividing by the total number of 
species as:
\begin{equation}
{\cal P}(R)={\cal N}\exp (-\frac{R^{2}}{2\rho ^{2}}) = {\cal N}\exp
\left( -\frac{(\ln n - \ln n_{0})^{2}}{2\sigma ^{2}}\right)\,,
\label{log_normal_1}
\end{equation}
where $\sigma = \rho \ln 2$ and where ${\cal N}$ is a normalization constant 
to be determined. So let us express the solution (\ref{Ps(n)_gamma}) as a 
function of $\ln n$ instead of $n$. After this change of variable a new, 
equivalent, probability distribution function arises, ${\cal P}(x)$, where 
$x=\ln n$, that has to satisfy $P_{s}(n)dn= {\cal P}_{s}(x)dx$, or, in other 
words,
\begin{equation}
{\cal P}_{s}(x)= \frac{dn}{dx}\,P_{s}(n) = n\,P_{s}(n)\,, 
\label{define}  
\end{equation}
which implies that 
\begin{equation}
\frac{d\ln {\cal P}_{s}(x)}{dx}=1+e^{x}\left[ -\psi (e^{x}+1)+
\psi (e^{x}+\lambda ^{*})-\psi (\nu ^{*}-e^{x})+\psi (N-e^{x}+1)\right]\,. 
\label{d_1}
\end{equation}
Setting equation (\ref{d_1}) equal to zero, and using $x=\ln n$, we obtain the 
position of maximum by finding the zero, $n_{0}$, of the equation 
\begin{displaymath}
\left[ \psi (n+1)-\psi (n+\lambda ^{*})+\psi (\nu ^{*}-n)-
\psi (N-n+1)\right] = \frac{1}{n}\,.
\end{displaymath}
In exactly the same way as for the Gaussian case, we can write the Taylor 
expansion up to second order: 
\begin{displaymath}
\ln {\cal P}_{s}(x)=\ln {\cal P}_{s}(x_{0})-\frac{1}
{2\sigma ^{2}}(x-x_{0})^{2}\,
\end{displaymath}
or, equivalently
\begin{displaymath}
{\cal P}_{s}(x)={\cal P}_{s}(x_{0})\exp (-\frac{(\ln n-\ln n_{0})^{2}}
{2\sigma ^{2}})
\end{displaymath}
where $x_{0}=\ln n_{0}$. Finally, using equation (\ref{define}), we get a 
lognormal expression for the mean field solution:
\begin{equation}
P_{s}(n)=\frac{K}{n}\exp (-\frac{(\ln n-\ln n_{0})^{2}}
{2\sigma ^{2}})
\label{log-normal}
\end{equation}
where $K=n_{0}P_{s}(n_{0})$.

The evaluation of the second derivative at $x=x_{0}$ allows us to fix a value 
for the variance $\sigma ^{2}$:
\begin{displaymath}
\frac{1}{\sigma ^{2}}=1-(n_{0})^{2}\left[ -\psi '(n_{0}+1)+\psi '(n_{0}+
\lambda ^{*})+\psi '(\nu ^{*}-n_{0})-\psi '(N-n_{0}+1)\right]\,. 
\end{displaymath}
Although in this case there is no way to derive a simple, yet sufficiently
general, analytical expression for the maximum $n_{0}$ and the variance 
$\sigma ^{2}$, in Fig \ref{fig_3} we have used the asymptotic series expansion 
for $\psi (z)$~\cite{abr} to calculate numerically both quantities. Once
again, since the lognormal distribution is completely specified by $n_0$
and $\sigma^{2}$, fixing these fixes the entire curve. The figure shows
that the lognormal approximation matches the exact solution well for 
intermediate to high immigration regimes. We also note that, in general,
the lognormal is a better fit to the exact solution than the Gaussian. 

Lognormal and logseries distributions have been used by ecologists to fit real
abundance data for years~\cite{Fisher43,Preston48,Preston62,i,eco,DeVries97}. 
Our results show that it is possible for both distributions to stem from the 
same general ecological process under different immigration regimes. If, for 
instance, we counted species abundances in a small area within a wood, the 
lognormal distribution would probably arise since that area is no doubt 
weakly isolated from the rest of the wood by external immigration. The same 
experiment performed in a rather isolated area might be expected to give rise 
to an empirical relative abundance distribution well fitted by a logseries 
function.

\begin{figure}
\centering
{\par\centering \resizebox*{0.8\columnwidth}{!}{\rotatebox{270}
{\includegraphics{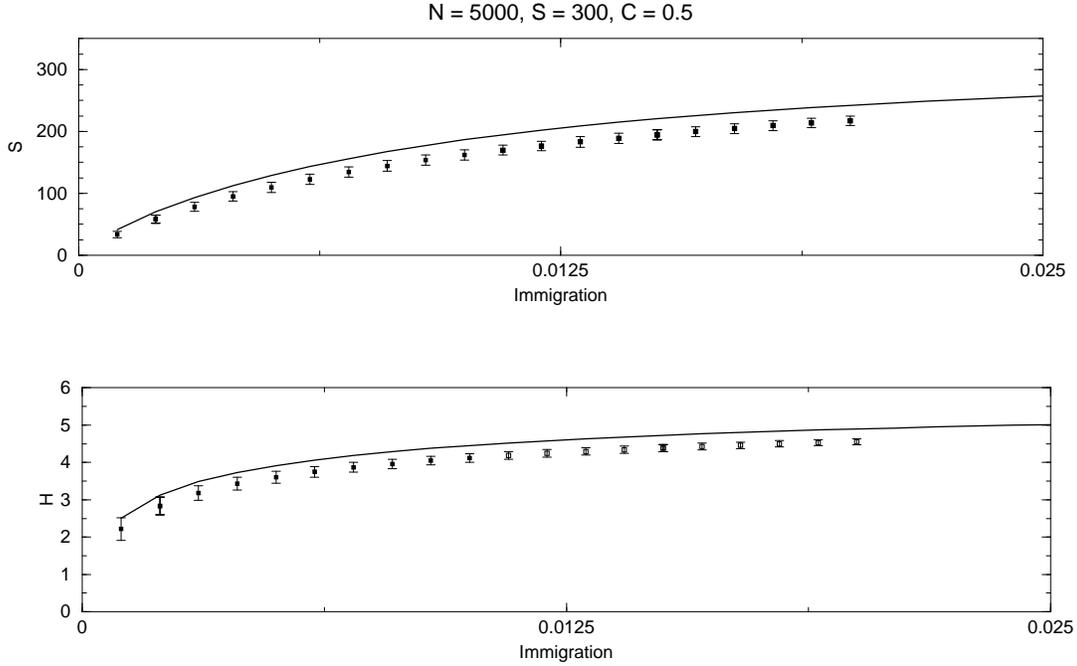}}} \par}
\caption{\label{Entropy}The expected value for the number of species and the 
Shannon entropy as a measure of diversity is computed at the steady state
using the stationary distribution (\ref{Ps(n)_beta}) for increasing values
of immigration parameter. Ensemble averages of these two quantities are also
shown for increasing discrete values of the same parameter. 1000 replicas of 
the model have been simulated. Values for the number of species and Shannon 
entropy have been recorded after 500000 simulation time steps, when the system 
has reached the stationary state.}
\end{figure}

Finally, in this section we calculate the diversity, $H$, of the ecosystem
(also called the Shannon entropy) for our model in the stationary state. It is 
defined by
\begin{equation}
H = - \sum_{i=1}^{S} p_{i}\ln p_{i}\, ,
\label{defdiv}
\end{equation}
where $p_i$ is the probability that an individual selected at random from the
system belongs to species $i$~\cite{eco}. Clearly, $p_{i} = n_{i}/N$, where 
$n_{i}$ is the number of individuals of species $i$ in the stationary state. 
On the other hand, within our mean field approach, the quantity which we can 
calculate is $P_{s}(n)$ --- the probability that a typical species will have 
$n$ individuals in the system when it is in the stationary state. Since the 
number of species with $n$ individuals in the system is just $SP_{s}(n)$, we 
may express (\ref{defdiv}) within our approximation as
\begin{equation}
H = - \sum_{n} SP_{s}(n)\,(n/N)\, \ln \left(n/N\right)\, .
\label{approxdiv}
\end{equation}
Multiplying (\ref{master}) by $n$ and summing, it is easy to find that 
$\langle n \rangle = N/S$ in the stationary state. Using this result we have
\begin{eqnarray}
H & = & - \frac{S}{N} \sum_{n} P_{s}(n)\, n \left( \ln n - \ln N \right)
\nonumber \\
& = & - \frac{S}{N} \left\{ \langle n\ln n \rangle - \langle n \rangle\, 
\ln N \right\} \nonumber \\
& = & \, \ln N \, - \, \frac{S}{N} \langle n\ln n \rangle\, . 
\label{div}
\end{eqnarray}
Therefore we only need to evaluate $\langle n\ln n \rangle$ in the stationary
state to find $H$. Fig \ref{Entropy} shows the result of performing this
evaluation using the stationary probability distribution $P_{s}(n)$, for 
increasing values of the immigration parameter $\mu$. For comparison purposes,
direct computation of the average number of species and the average entropy 
for 1000 replicas of the model and its standard deviation is shown. It can be 
seen that for relatively low immigration rates the system tends to be 
saturated admitting as many species as possible. As immigration increases, the 
Shannon entropy grows steeper than the number of species does, meaning that 
immigration tends to equalize the number of individuals of different species 
first, rather than increase the actual number of species in the system.

In Figs \ref{Conn} and \ref{Entropy} it can be seen that ensemble average
curves for the expected number of species in the stationary state deviate
systematically from the mean field approximation that we have implemented 
through the master equation (\ref{master}), even though they show the same
qualitative behaviour. The explanation for this slight disagreement comes 
from the way we are estimating the probability of an effective interaction 
within the system. When transition probabilities are discussed (Eqns.
(\ref{gn}) and (\ref{rn})), the probability of interaction is split into a 
product of the probability of picking two potential interacting individuals 
multiplied by the probability of having an actual link between the species 
they belong to. However, these two events are not independent. The assumption 
of independence is just an approximation that allows us to simplify the system 
and get some insight into the dynamical processes that take place during 
simulations. In particular, for any parameter choice, the individual based 
model has more interactions than expected and cannot maintain the same number 
of species as predicted by our mean field approximation.  
 
\section{time dependence}

What can our model say about the assembly of an ecological community? Whenever
species colonize a new island or any empty space, a new community builds
up from scratch. The process that takes place is called succession by
ecologists. The assembly of an ecological community has been studied both from 
the theoretical and empirical point of view. Many patterns have been found
during the process of ecological succession (see~\cite{Margalef97} 
for a review). For instance, the number of 
species grows in a particular way that depends on the immigration 
from a biogeographical species pool. If our model is to make 
any prediction about succession, simulation time must have a direct 
meaning in terms of physical time. In our model the only connection to 
physical time comes from the immigration parameter, but the model has in fact 
two time scales. The external one is defined by the flux of individuals from 
the biogeographical pool and the internal one is defined by the flux of
individuals (birth-death process) as a  consequence of the internal dynamics
(pairwise random encounters) within the system. Our immigration parameter
captures the relative importance of these two different temporal processes.  

Therefore, having investigated the properties of the stationary state in the
last section, we now move on to the study of the time evolution of the system
within the mean-field approximation. The master equation (\ref{master}) has 
transition probabilities $g_n$ and $r_n$ which are non-linear in $n$, so that 
an exact solution for the time-dependent behavior is not possible. However, 
since in the problem of interest $N$ is very large, the possibility of 
performing a large-$N$ analysis suggests itself. In this section we will 
describe the application of such an analysis --- specifically van Kampen's 
large-$N$ method~\cite{van} --- to our model. This method has a number of 
attractive features, for instance, the macroscopic (i.e. deterministic) 
equation emerges naturally from the stochastic equation as a leading order 
effect in $N$, with the next to leading order giving the Gaussian broadening 
of $P(n,t)$ about this average motion. The method is clearly presented 
in~\cite{van}, so we will only give a brief outline of the general idea and 
stress the application to the model of interest in this paper. 

If we take the initial condition on (\ref{master}) to be 
$P(n,0) = \delta_{n,m}$, we would expect, at early times at least, $P(n,t)$ to
have a sharp peak at some value of $n$ (of order $N$), with a width of order
$N.N^{-1/2} = N^{1/2}$. It is therefore, natural to transform from the
stochastic variable $n$ to the stochastic variable $\xi$ by writing
\begin{equation}
n = N \phi (t) + N^{1/2} \xi
\label{ntoxi}
\end{equation}
where $\phi (t)$ is some unknown macroscopic function which will have to be
chosen to follow the peak in time. A new probability distribution $\Pi$ is 
defined by $P(n,t) = \Pi (\xi ,t)$, which implies that
\begin{equation}
\dot{P} = \frac{\partial \Pi}{\partial t} - N^{1/2}\,\frac{d\phi}
{dt}\,\frac{\partial \Pi}{\partial \xi}\, .
\label{transform}
\end{equation}
The master equation (\ref{master}) may be written 
\begin{equation}
\dot{P}_{n} = \left( {\cal E} - 1 \right) r_{n}P_{n} + 
\left( {\cal E}^{-1} - 1 \right) g_{n}P_{n}\, ,
\label{altmaster}                             
\end{equation}
where ${\cal E}$ (${\cal E}^{-1}$) is an operator which changes $n$ into 
$n + 1$ ($n - 1$), e.g. if $f_n$ is an arbitrary function of $n$, then 
${\cal E}f_n = f_{n + 1}$. In terms of $\xi$:
\begin{equation}
{\cal E}^{\pm 1} = 1 \pm N^{-1/2}\, \frac{\partial\ }{\partial \xi} +  
\frac{1}{2\,!}\, N^{-1}\, \frac{\partial^{2}}{\partial \xi^{2}} + \ldots\, .
\label{operators}
\end{equation}
Using (\ref{ntoxi})-(\ref{operators}) the original master equation for $P(n,t)$ 
can be rewritten as an equation for $\Pi(\xi ,t)$. By rescaling the time
according to $\tau = t/N$, a hierarchy of equations can be derived by
identifying terms order by order in powers of $N^{-1/2}$. The first two of
these are:
\begin{equation}
\frac{d\phi}{d\tau} = \alpha_{1,0}(\phi)
\label{macroscopic}
\end{equation}
and
\begin{equation}
\frac{\partial \Pi}{\partial \tau} = - \alpha'_{1,0}(\phi)\,
\frac{\partial\ }{\partial \xi}\left( \xi \Pi \right) +  
\frac{1}{2}\, \alpha_{2,0}(\phi)\, \frac{\partial^{2}\Pi}{\partial \xi^{2}}\, , 
\label{FP}
\end{equation}
where
\begin{eqnarray}
\alpha_{1,0}(\phi) & = & \frac{\mu}{S} - \mu \phi \nonumber \\
\alpha_{2,0}(\phi) & = & 2C^{*}(1 - \mu )\phi (1 - \phi ) + \frac{\mu}{S} 
+ \frac{\mu}{S}\, (S - 2)\phi\, . 
\label{alphas}
\end{eqnarray}
The first equation (\ref{macroscopic}) is the macroscopic equation for 
$\phi (\tau)$. It is easily solved to give
\begin{equation}
\phi(\tau) = \phi(0)e^{-\mu \tau} + \frac{1}{S}\, (1 - e^{-\mu \tau})\, .
\label{phi_tau}
\end{equation}
Initially we ask that $\xi(0) = 0$, which means that $\phi(0) = n(0)/N = m/N$.
Going back to the $t$ variable gives
\begin{equation}
\phi(t) = \frac{m}{N}\, e^{-\mu t/N} + \frac{1}{S}\, (1 - e^{-\mu t/N})\, .
\label{phi_t}
\end{equation}
The second equation (\ref{FP}) is a linear Fokker-Planck equation whose
coefficients depend on time through $\phi$ given by (\ref{phi_t}). It is 
straightforward to show that the solution to this equation is  a Gaussian and so
it is only necessary to determine $\langle \xi \rangle_{\tau}$ and
$\langle \xi^{2} \rangle_{\tau}$ to completely characterize $\Pi(\xi ,t)$. By
multiplying (\ref{FP}) by $\xi$ and $\xi^{2}$ and using integration by parts,
one finds~\cite{van}
\begin{eqnarray}
\partial_{\tau}\langle \xi \rangle_{\tau} & = & \alpha'_{1,0}(\phi) 
\langle \xi \rangle_{\tau} \ \ \ {\rm and\ } \nonumber \\
\partial_{\tau}\langle \xi^{2} \rangle_{\tau} & = & 2\alpha'_{1,0}(\phi) 
\langle \xi^{2} \rangle_{\tau} + \alpha_{2,0}(\phi)\, .
\label{meanandvar}
\end{eqnarray}
In our case 
$\partial_{\tau}\langle \xi \rangle_{\tau} = -\mu \langle \xi \rangle_{\tau}$ 
and so
\begin{equation}
\langle \xi \rangle_{\tau} = \langle \xi \rangle_{0}\, e^{-\mu \tau} = 0\, ,
\label{mean}
\end{equation}
since we have already assumed that $\langle \xi \rangle_{0} = 0$. A
straightforward, but tedious, calculation now gives
\begin{eqnarray}
\langle \xi^{2} \rangle_{\tau} & = & \frac{(\eta + \mu)}{\mu}\, 
\frac{(S - 1)}{S^{2}}\, \left[ 1 - e^{-2\mu \tau} \right] \nonumber \\
& + & {\cal A} \frac{(2\eta + \mu)}{\mu}\, \frac{(S - 2)}{S}\, e^{-\mu \tau}
\left[ 1 - e^{-\mu \tau} \right] \nonumber \\
& - & 2\eta {\cal A}^{2}\tau e^{-2\mu \tau}\, , 
\label{var}
\end{eqnarray}
where $\eta = C^{*}(1 - \mu )$ and ${\cal A} = (m/N) - (1/S)$. 

\begin{figure}
\centering
{\par\centering \resizebox*{1\columnwidth}{!}{\rotatebox{270}
{\includegraphics{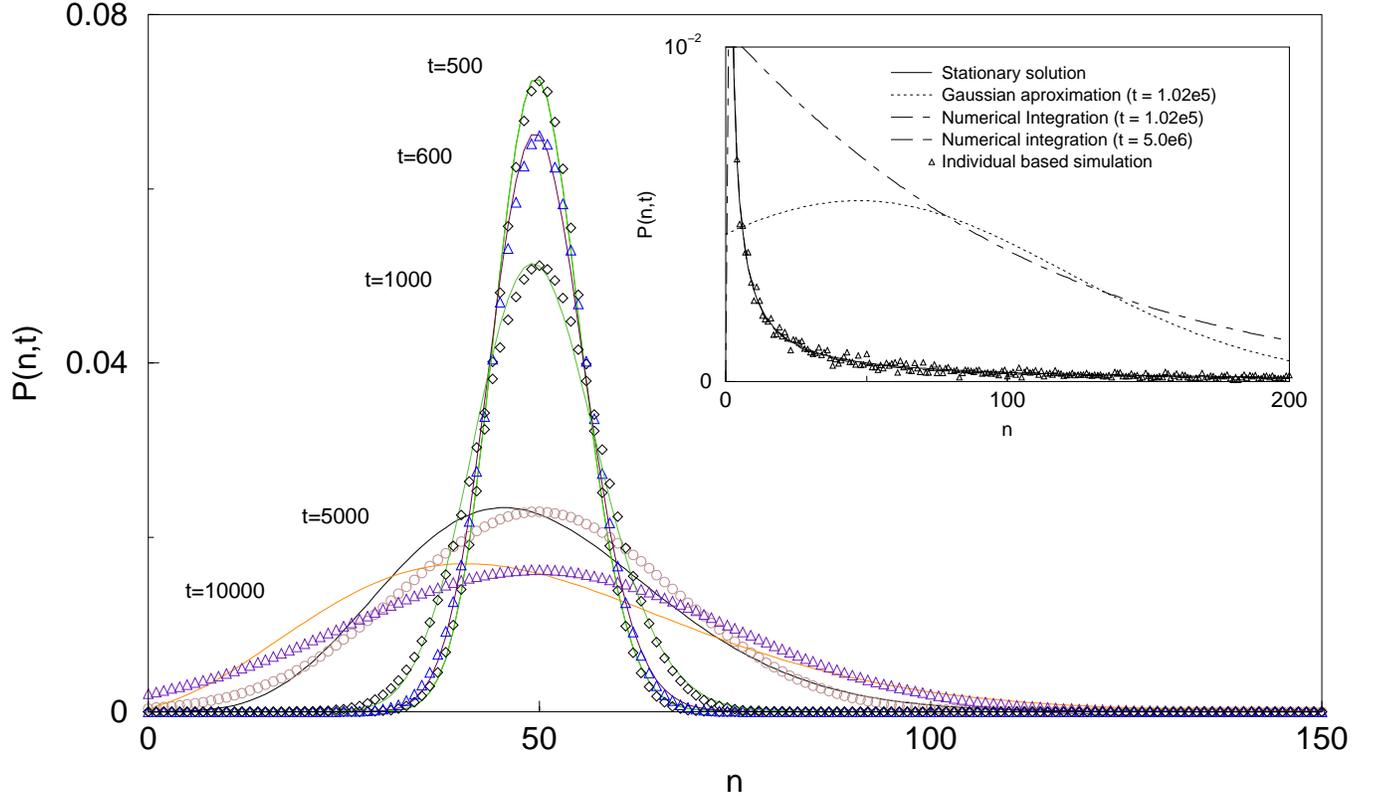}}} \par}
\caption{\label{Gaussian1} Temporal evolution of the relative abundance 
distribution $P(n,t)$. The parameter values are $N=1000$, $S=50$, $C=0.4$ and 
$\mu =0.001$. An individual-based simulation, the numerical integration of 
$P(n,t)$ at two successive times, and the exact stationary solution are also 
presented (inset).}
\end{figure}

\begin{figure}
\centering
{\par\centering \resizebox*{1\columnwidth}{!}{\rotatebox{270}
{\includegraphics{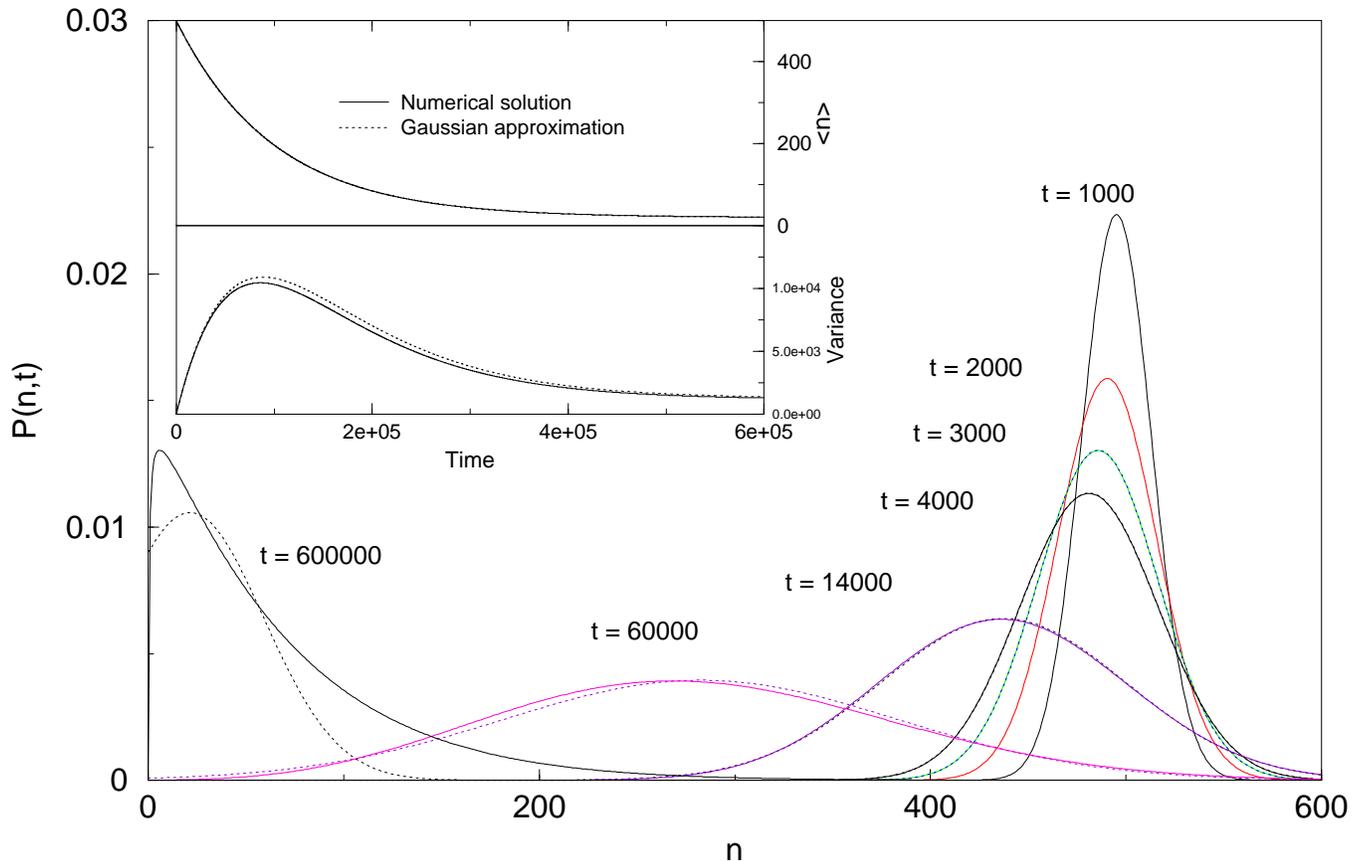}}} \par}
\caption{\label{Gaussian2} Temporal evolution of the relative abundance 
distribution $P(n,t)$. The parameter values are $N=1000$, $S=50$, $C=0.4$ and 
$\mu =0.01$. The temporal evolution of the variance and the average abundance 
computed using (\ref{var}) and (\ref{phi_t}) respectively and from the master 
equation is also shown (inset).}
\end{figure}

We have already commented that the solution of (\ref{FP}) is a Gaussian. 
Specifically
\begin{equation}
P(n,t)=\frac{1}{\sqrt{2\pi N\, \langle \xi ^{2}\rangle _{\tau}}}
\exp\left( -\frac{(n-N\, \phi (t))^{2}}{2 N\,  
\langle \xi ^{2}\rangle _{\tau}}\right)\,, 
\label{Large-N}
\end{equation}
where $\langle \xi ^{2}\rangle _{\tau}$ and $\phi (t)$ are given by equations
(\ref{var}) and (\ref{phi_t}) respectively. In Fig \ref{Gaussian1} and
\ref{Gaussian2} a comparison between the numerical integration of the master
equation and the Gaussian solution for different times is shown. The Gaussian
behavior is lost for large times. In figure \ref{Gaussian2} the Gaussian
behavior is maintained longer due to a higher immigration rate; as the
immigration rates increase still further, the Gaussian form persists for even
larger times. 

In order to compare the time behavior of the mean field approach introduced
in this work through the master equation (\ref{master}) with the time 
behavior of the individual based model (IBM) defined by the rules presented in 
section I, one should carefully define what is meant by time. Individual based 
simulations are performed by the iteration of an algorithm from the first 
step up to a given number of updating steps. In section 1, such an updating 
step has been defined as a time unit. Let us call it the simulation time unit. 
In~\cite{I} a different operational choice was made. Whatever the convention 
is, a clear distinction must be made between the simulation time and the 
physical time needed to compare simulation results with either 
the numerical integration of the master 
equation (\ref{master}) or the large-$N$ solution derived in this section. 
The question then arises: how is physical time to be tracked in any stochastic 
realization of the IBM? To analyze this point we will follow an argument given 
by Renshaw~\cite{Renshaw93}. At any time $t$, the probability of an event 
occurring in the system can be estimated. Such a probability depends on the 
system 
configuration, i.e, the abundance of all present species, and on the relative 
immigration rate $\mu$ in relation to the internal dynamics rate $1-\mu$. Both 
rates have dimensions of $T^{-1}$. Obviously, it depends also on the other 
parameters of the model ($N$, $S$ and $C$). Although the method described in 
Ref \cite{Renshaw93} estimates every transition probability rate
for all possible events in the system, there is no need to estimate the 
probability of this rather high number of possible events. 
There are only two relevant temporal processes: immigration and internal  
dynamics. So, it is enough to consider these two different possibilities: 

\begin{enumerate}
\item An immigration event $I$ occurs if any species from the pool happens
to enter the system. The probability of a pool species entering the system in
any small $dt$ is: 
\begin{displaymath}
Pr\{I\}=r_{I}(t)\, dt\,,
\end{displaymath}
where the immigration rate is
\begin{displaymath}
r_{I}=\frac{\mu }{S}\sum _{i=1}^{S}\left( 1-\frac{n_{i}(t)}{N}\right)\,.
\end{displaymath}
\item An internal dynamics event $D$ occurs when the interaction between a
pair of individuals from two potentially interacting species gives rise to a
change in their abundances. The probability of such an event occurring in any
small $dt$ can be written as:
\begin{displaymath}
Pr\{D\}=r_{D}\, dt\,,
\end{displaymath}
where the internal dynamics rate is
\begin{displaymath}
r_{D}=(1-\mu )\, \sum _{i=1}^{S}\sum _{j\in 
{\cal G}(i)}\frac{n_{i}(t)}{N}\frac{n_{j}(t)}{N}\,,
\end{displaymath}
and where ${\cal G}(i)$  must be understood 
as the set of species --- different from $i$ --- that are connected to $i$ 
through the pre-defined interaction matrix $\bf \Omega$. 
\end{enumerate}
Since the two events defined are independent from each other, the probability
of occurrence of any one of them in any small $dt$ is:
\begin{displaymath}
Pr\{I\: \cup \: D\}=(r_{I}+r_{D})\, dt\,.
\end{displaymath}
When an event occurs there is a change in the actual configuration of the 
system either by immigration or by internal dynamics and the rates must be 
calculated again. So, approximately, on average the number of such effective 
events in any time interval of length $t$ would be $(r_{I}+r_{D})\,t$, and 
would be distributed as a Poisson random variable with that mean. The 
important point is that now the probability of having no events in any time 
interval of length $t$, i.e, for any time between $0$ and $t$, can be written 
as:
\begin{equation}
Pr\{0\}=e^{-(r_{I}+r_{D})\, t}\,.
\label{exp}
\end{equation}
According to (\ref{exp}), the probability of having at least one event is 
$1-e^{-(r_{I}+r_{D})\,t}$ --- the cumulative probability distribution for 
an exponentially distributed random variable. Therefore, the time to the next 
event is an exponentially distributed random variable with expectation 
$1/(r_{I}+r_{D})$. Then, we should sample that distribution in order to 
predict when the next effective event will take place. Accumulating these 
inter-event times during simulations we are able to track the physical time, 
which have the same units as $[r_{I}+r_{D}]^{-1}$, so the same time units 
which arise in the master equation (\ref{master}).

\begin{figure}
\centering
{\par\centering \resizebox*{0.9\columnwidth}{!}{\rotatebox{270}
{\includegraphics{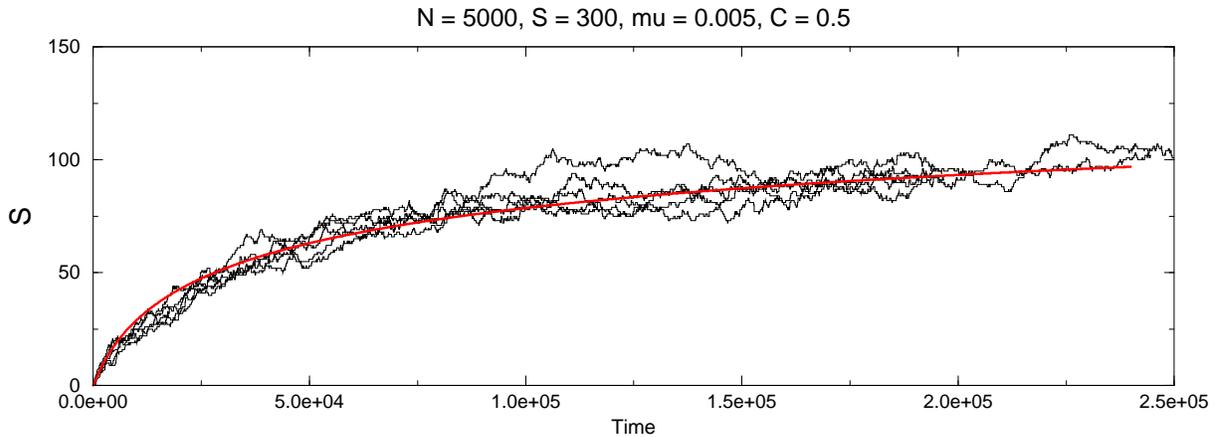}}} \par}
\caption{\label{Time_evo} Time evolution of the number of species in the 
system. The individual-based simulations match the expected value from the 
master equation. The system starts with no species. During the stochastic 
assembly process the number of species fluctuates, but the average behaviour is 
captured by the mean field approach represented by the master equation 
(\ref{master}).}
\end{figure}

In figure \ref{Time_evo}, the time evolution of the number of species in the
system is shown. Different stochastic realizations of the IBM are presented.
The numerical integration of the master equation allows the estimation of the
expected number of species at any time in the system $<S>$ through 
(\ref{defavs}). The average behavior of the different stochastic simulations 
is well captured by the prediction given by (\ref{defavs}) where $P(0,t)$ is 
computed at each numerical integration time step. 

\begin{figure}
{\par\centering \resizebox*{0.7\columnwidth}{!}{\rotatebox{270}
{\includegraphics{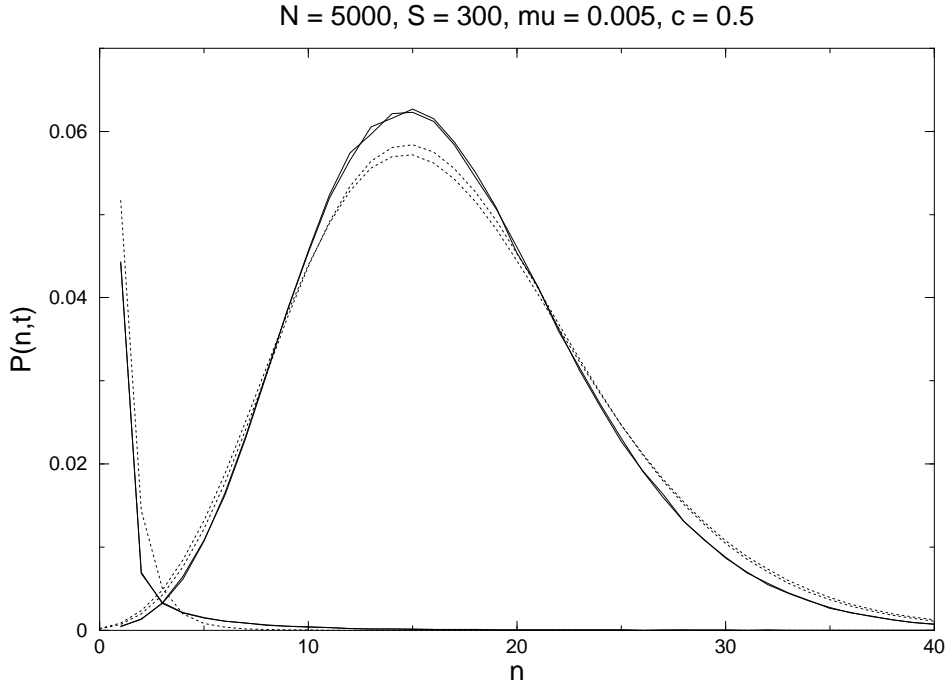}}} \par}
\caption{\label{Tempo}Individual-based model and the temporal evolution of 
$P(n,t)$. Comparison between the mean field approach (dotted line) and a 2000 
ensemble average after 5000 simulation time steps for random mixture and empty 
system initial conditions. In the first case, two different numerical 
integrations have been performed for $t=9946$ and $t=10400$. In the second 
case, just one numerical integration has been performed until $t=6725$. The 
reason for these different times is that 5000 simulation updating steps can 
represent more or less physical time depending on the initial configuration of 
the system.}
\end{figure}

In Fig \ref{Tempo} the probability of having a species represented by $n$
individuals at particular early times, $P(n,t)$, has been plotted. It has
been computed by performing a numerical integration of the master equation 
(\ref{master}) (dotted line) and by means an ensemble average for the
individual-based model after 5000 simulation time steps. Two extremely 
different initial conditions have been used. In the first one, there are no 
species in the system at time $0$. Species enter the system and either 
establish themselves in it or not, performing what could be
called a stochastic community assembly. In the second initial state, all
species are represented in approximately equal numbers.  
Obviously, the one-humped distributions are obtained when the initial condition
is a random mixture of species, which is represented by $P(n,0)=1$ 
if $n=N/S$ and $P(n,t)=0$ if $n\neq N/S$ in the master equation approach. 
The purely decreasing distributions are obtained when the initial state  
is a completely empty system. The agreement between the mean field approach 
represented by the master equation and the simulations is seen to be
reasonable.  

As it has been shown in Figs \ref{Gaussian1} and \ref{Gaussian2},  eventually, 
the probability distribution deviates from a Gaussian. 
While it is true that one could in principle calculate these 
non-Gaussian effects using van Kampen's approach (by taking higher order terms 
in $N^{-1/2}$ into account), the method gets increasingly cumbersome. 
Therefore, in the next section, we adopt a totally different approach to the 
calculation of time-dependence, which is able to give information about 
$P(n,t)$ at late times.

\section{generating function}

The technique we will use to probe the time-dependence of $P(n,t)$ in this
section is based on the solution of the differential equation satisfied by the
generating function
\begin{equation}
F(z,t) = \sum^{N}_{n=0}\, P(n,t) z^{n}
\label{defgenfun}
\end{equation}
for our model in the mean-field approximation. Starting from (\ref{master}) 
the derivation of this equation proceeds along standard lines~\cite{van,gar} to
yield
\begin{equation}
\frac{\partial F}{\partial s} = -z(1 - z)^{2} 
\frac{\partial^{2}F}{\partial z^{2}} + (1 - z) (\alpha + \beta z)
\frac{\partial F}{\partial z} - \gamma (1 - z)F
\label{pde}
\end{equation}
where we have introduced a new time
\begin{equation}
s = \zeta t\, , \ \ \ {\rm where} \ \ \zeta = \frac{C^{*}}{N(N - 1)}\, ,
\label{newtime}
\end{equation}
and where the constants $\alpha$, $\beta$ and $\gamma$ are defined by
\begin{displaymath}
\alpha = \nu^{*} - 1 = \lambda^{*} (S - 1) + N - 1\, ,
\end{displaymath}
\begin{equation}
\beta = \lambda^{*} + 1 - N\ \ \ {\rm and} \ \ \ \gamma = N\lambda^{*}\, . 
\label{aibega}
\end{equation}
The conditions on $F$ are
\begin{equation}
F(1,t) = 1 \ \ \ {\rm and} \ \ \ F(z,0) = z^{m}\, ,
\label{conds_F}
\end{equation}
and follow from the normalization condition $\sum_n P(n,t) = 1$ and the
initial condition $P(n,0) = \delta_{n,m}$ respectively.

The partial differential equation (\ref{pde}) is separable: if we write 
$F(z,s) = {\cal S}(s) \Phi(z)$, then ${\cal S}(s) = e^{-\lambda s}$, where 
$\lambda$ is a constant. The equation for $\Phi$ is then
\begin{equation}
z(1 - z)^{2}\frac{d^{2}\Phi}{dz^{2}} - 
(1 - z) (\alpha + \beta z) \frac{d\Phi}{dz} + 
\gamma (1 - z)\Phi = \lambda \Phi\, .
\label{ode}
\end{equation}
This can be brought into a more standard form by the change of variables
\begin{equation}
\Phi = (1 - z)^{N}\phi \ \ \ {\rm and} \ \ u = \frac{1}{1 - z}\, .
\label{newvar}
\end{equation}
The new form of the equation is 
\begin{equation}
u(1 - u)\frac{d^{2}\phi}{du^{2}} +
\left[ c - (a + b + 1)u \right] \frac{d\phi}{du} - 
a b \phi = 0\, ,
\label{hyper}
\end{equation}
where
\begin{displaymath}
a + b = 1 - \lambda^{*} - N - \nu^{*}\, ,  
\end{displaymath}
\begin{equation}
ab = N\left( \lambda^{*} + \nu^{*} - 1 \right) - \lambda\ \ \ {\rm and} 
\ \ \ c = 1 - \lambda^{*} - N\, .
\label{abc}
\end{equation}
The reason for making the transformation (\ref{newvar}) is that (\ref{hyper})
is the standard form for the hypergeometric equation~\cite{abr}, which has the
two independent solutions:
\begin{eqnarray}
\phi^{(1)}_{\lambda} & = & u^{-a}\, F(a, a - c + 1, a - b + 1; u^{-1})\ \ \
{\rm and}\nonumber \\
\phi^{(2)}_{\lambda} & = & u^{-b}\, F(b, b - c + 1, b - a + 1; u^{-1})\, .
\label{phi_solns}
\end{eqnarray}
Now $u^{-1} = 1 - z$ and so in terms of the original variables
\begin{eqnarray}
\Phi^{(1)}_{\lambda} & = & (1 - z)^{N+a}\, F(a, a - c + 1, a - b + 1; 1 - z)\ 
\ \ {\rm and}\nonumber \\
\Phi^{(2)}_{\lambda} & = & (1 - z)^{N+b}\, F(b, b - c + 1, b - a + 1; 1 - z)\, .
\label{Phi_solns}
\end{eqnarray}
The general solution to (\ref{pde}) is then 
\begin{equation}
F(z,t) = \sum_{\lambda} \left\{ v_{\lambda}\,\Phi^{(1)}_{\lambda} +
w_{\lambda}\,\Phi^{(2)}_{\lambda} \right\}\, e^{-\lambda\zeta t}\, ,
\label{gensoln}
\end{equation}
where $\{v_{\lambda}\}$ and $\{w_{\lambda}\}$ are sets of arbitrary constants.

To determine the arbitrary constants in (\ref{gensoln}), the conditions
(\ref{conds_F}) have to be implemented. The details are given in Appendix B,
where it is shown that the required solution is
\begin{equation}
F(z,t) = \sum_{k=0}^{N}\, w_{k}\, (1 - z)^{k}\,
F(k - N,k + \lambda^{*},2k + \lambda^{*}S; 1 - z)\, 
e^{-k(k + \lambda^{*}S - 1) \zeta t}\, ,
\label{final_form}
\end{equation}
where the constants $\{w_{k}\}$ are determined by
\begin{equation}
\sum^{n}_{k=0}\, w_{k}\, (-1)^{n-k}\, {N-k \choose n-k}\, 
\frac{\Gamma(n + \lambda^{*})\, \Gamma(2k + \lambda^{*}S)}
{\Gamma(k + \lambda^{*})\, \Gamma(n + k + \lambda^{*}S)} =
\left\{ \begin{array}{ll} 
(-1)^{n}\,{m \choose n}, & \mbox{\ if $n \le m$} \\ 
\ \ \ \ \ \ 0 \ \ \ \ \ \ , & \mbox{\ if $n > m$\, .}
\end{array} \right.
\label{finding_wk}
\end{equation}
This equation holds for all allowed values of $n$ ($n = 0,1,\ldots,N$). We
therefore have $(N + 1)$ linear conditions for the $(N + 1)$ constants $w_k$
($k = 0,1,\ldots,N$), and so can determine them uniquely. Thus,
(\ref{final_form}) together with (\ref{finding_wk})  provide a complete
solution to the partial differential equation (\ref{pde}). 

Although the solution is not in closed form, it is possible to obtain the $w_k$
for small values of $k$ rather easily. For $n=0$, (\ref{finding_wk}) involves
only $w_0$, for $n=1$ only $w_0$ and $w_1$, and so on. The expressions for the
first three constants are
\begin{eqnarray}
w_{0} & = & 1\, , \ \ \ w_{1} = \frac{N}{S} - m\, , \nonumber \\
w_{2} & = & \frac{(N - 1)(1 + \lambda^{*})}{(2 + \lambda^{*}S)}\, 
\left[ \frac{N}{S} - m \right] \ - \ \frac{N(N - 1)}{2S}\,
\frac{(1 + \lambda^{*})}{(1 + \lambda^{*}S)}\ + \ \frac{m(m - 1)}{2}\, ,
\label{w012}
\end{eqnarray}
the result for $w_0$ confirming what we already knew. These results are very 
useful because, as is clear from (\ref{final_form}), the large-time behavior
of the system is governed by small values of $k$. In this case, as we will now 
show, an explicit form for $P(n,t)$ can be found.

\begin{figure}
{\par\centering \resizebox*{1\columnwidth}{!}{\rotatebox{270}{\includegraphics
{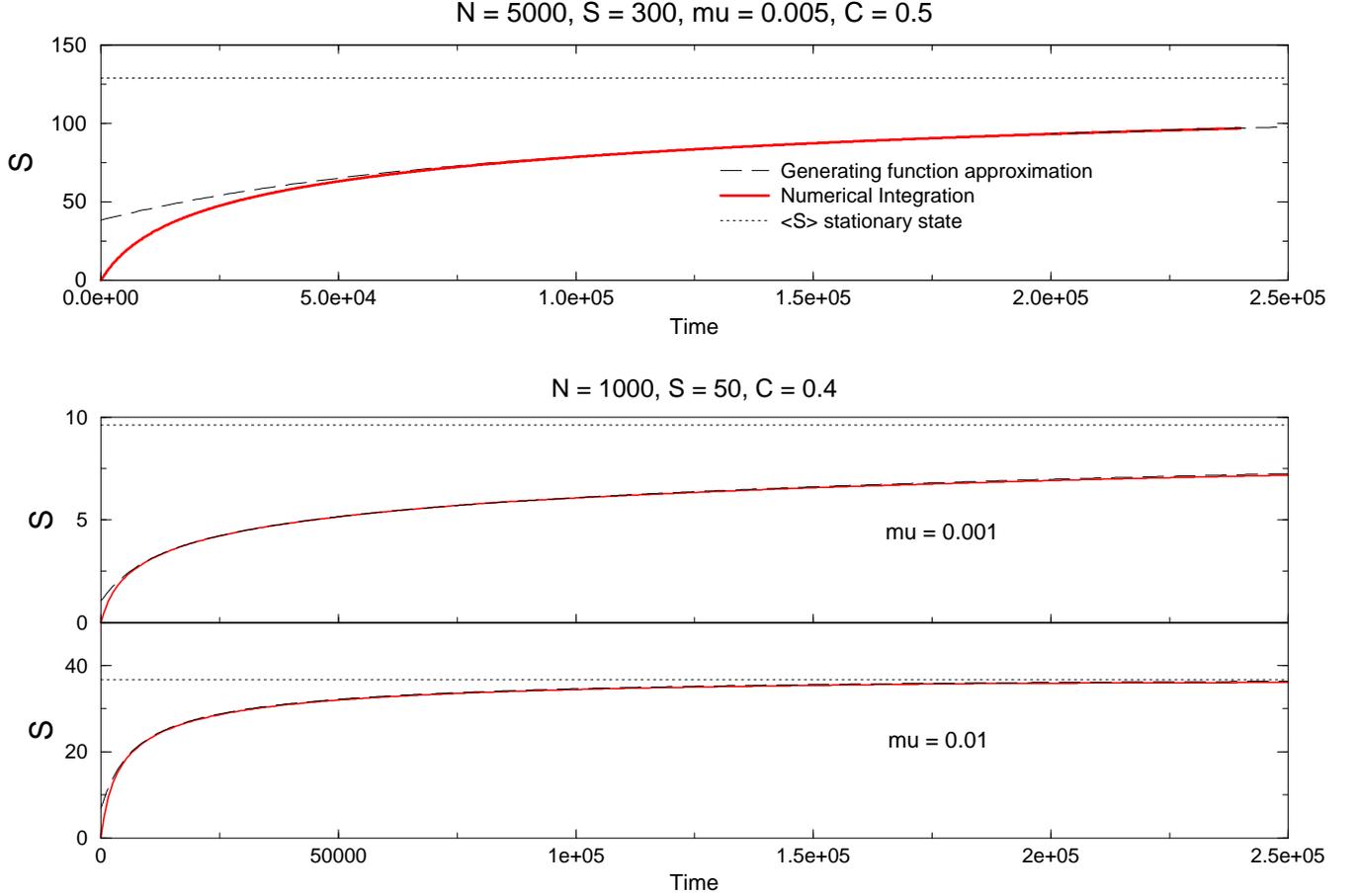}}} \par}
\caption{\label{fig19} Temporal evolution of the expected value for the number 
of species in the system. A truncated solution provided by (\ref{P(0,t)}) 
is shown (dashed-lines). Only the first 20 $w_{k}\protect$  coefficients have 
been computed. The agreement with the numerical integration of the
master equation (bold lines) gets better and better the larger time is. For
$N=1000$, $S=50\protect$, and  $C=0.4$ such an agreement is quite good even for
early times.}
\end{figure}

To find $P(n,t)$ we have the identify the coefficient of $z^{n}$ in 
(\ref{final_form}). In Appendix B it is shown that this leads to
\begin{equation}
P(n,t) = \sum^{N}_{k=0}\, w_{k}\,\chi(n,k)\, 
e^{-k(k + \lambda^{*}S -1)\zeta t}\, ,
\label{final_result}
\end{equation}
where 
\begin{eqnarray}
\chi(n,k) & = & \sum^{r={\rm min}\{N-k,n\}}_{r={\rm max}\{n-k,0\}}\, 
(-1)^{n-r}\, {k \choose n-r}\, {N-k \choose r} \nonumber \\ \nonumber \\
& \times & {\Gamma (r + k + \lambda^*) \over \Gamma(k + \lambda^*)} 
{\Gamma (\nu^* - r) \over \Gamma(k + \nu^* - N)} 
{\Gamma (2k + \lambda^* + \nu^* - N) \over \Gamma(k + \lambda^* + \nu^*)}\, .
\label{full_chi}
\end{eqnarray}
This result appears to be rather complicated, but fortunately it simplifies in
many cases of interest. For instance, suppose we wish to find $P(0,t)$: the
time-evolution of the probability that there are no individuals of the species 
present in the system. Since $\chi(0,k)$ has only one term in the sum 
($r = 0$),
\begin{eqnarray} 
P(0,t) & = & \Gamma(\nu^{*})\,\sum^{N}_{k=0}\,w_{k}
\frac{\Gamma(2k + \nu^{*} + \lambda^{*} - N)}
{\Gamma(k + \nu^{*} - N)\,
\Gamma(k + \nu^{*} + \lambda^{*})}\, e^{-k(k + \lambda^{*}S - 1)\zeta t}
\nonumber \\ \nonumber \\
& = & P_{s}(0) \left\{ 1 + w_{1} \frac{S (\lambda^{*}S + 1)}
{(S - 1)(\lambda^{*}S + N)}\, e^{-\lambda^{*}S \zeta t} \right. \nonumber \\
\nonumber \\
& + & \left. w_{2} \frac{S (\lambda^{*}S + 1) (\lambda^{*}S + 2)
(\lambda^{*}S + 3)} 
{(S - 1)(\lambda^{*}(S - 1) + 1)(\lambda^{*}S + N) (\lambda^{*}S + N + 1)}\,
e^{-2(\lambda^{*}S + 1)\zeta t} + \ldots \right\}\, ,
\label{P(0,t)}
\end{eqnarray}
which describes the approach to the stationary state at large times. In Fig
\ref{fig19} the temporal evolution of the expected number of species, estimated
as \( S(t)=S(1-P(0,t)) \) is shown. The temporal solution provided by this
method is exact as long as the total number of coefficients can be computed 
without numerical error. In Fig \ref{fig19} a truncated, approximated solution 
is compared with the straightforward numerical integration of the master 
equation. Complete agreement is observed at large times. 

If $n \neq 0$, the large time behavior of $P(n,t)$ still has a relatively simple
form:
\begin{eqnarray} 
P(n,t) & = & P_{s}(n) + w_{1}\,\chi(n,1)\, 
e^{-\lambda^{*}S \zeta t} \nonumber \\ 
& + & w_{2}\,\chi(n,2)\, e^{-2(\lambda^{*}S + 1)\zeta t} + 
\ldots \, ,
\label{large_t}
\end{eqnarray}
where $\chi(n,1)$ has only two terms, $r = n - 1$ and $r = n$ (only one if $n =
0$ or $n = N$) and $\chi(n,2)$ has only three terms, $r = n - 2, n - 1$ and 
$n$ (fewer if $n = 0, 1, N - 1$ or $N$). 

\begin{figure}
{\par\centering \resizebox*{1\columnwidth}{!}{\rotatebox{270}{\includegraphics
{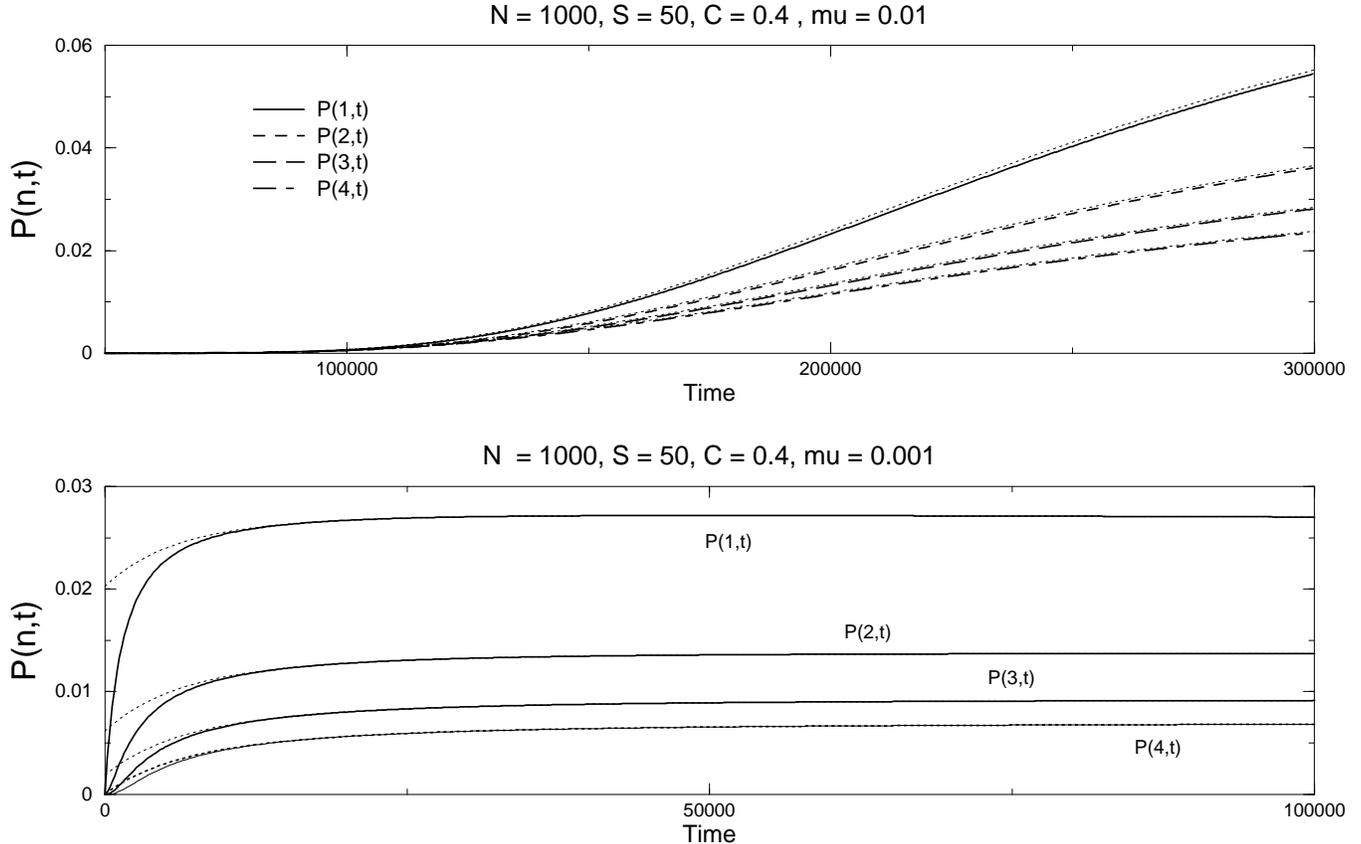}}} \par}
\caption{\label{fig20} Temporal evolution of probability of having any species
represented by 1, 2, 3, and 4 individuals computed directly from 
(\ref{final_result}) in dotted lines, where again a truncated solution is 
used (only the first 20 $w_{k}\protect$ are considered), compared with the
numerical integration of the master equation (bold lines). The initial
condition is $P(n,0)=\delta_{m,n}$ where $m=500$ in the upper plot and $m=0$ 
in the lower one.}
\end{figure}

In Fig \ref{fig20}, the computation of $P(n,t)$ for $n=1,2,3,4$ is shown. The
solution is approximate because again just the first 20 $w_{k}$ 
have been considered, although (\ref{final_result}) would be an exact 
solution as long as all of the terms from $k=0$ to $k=N$ could be
summed without numerical error. For practical reasons this is obviously
not possible. In particular, at early times, the truncation of
(\ref{final_result}) introduces errors in $P(n,t)$. The same is true when $n$
is too large, because the sums in (\ref{finding_wk}) and (\ref{full_chi}) 
are too long to be computed without errors and some numerical instabilities 
arise. 

\section{Conclusions}

In this paper we have analysed a model which has a structure which is
rich enough to show many of the underlying patterns seen in real ecosystems, 
but is still sufficiently simple for a variant of the mean field approximation 
to be applied to obtain analytical results. The most straightforward question 
that can be asked concerns the nature of the stationary state. Within the mean 
field approximation an exact form for this stationary distribution may be 
derived. We found that this exact result reduced to the logseries distribution 
in the regime of low immigration and to the lognormal distribution in the 
regime of moderate to high immigration. These two distributions have been 
discussed by ecologists for decades as possible forms for the species 
abundance distributions. Our approach gives a clear interpretation to the 
parameters on which they depend. This fact has practical consequences for 
conservation biology in order to determine the potential richness ($S$), the 
global size ($N$) and the degree of isolation (${\mu}$) of a community. We have
therefore shown how logseries and lognormal distributions can arise as two 
different limits of a single distribution, a distribution which is, moreover, 
the stationary distribution of a well-motivated ecological model. We also found 
evidence for the hyperbolic relation between the connectivity and the average 
number of species --- the so-called $C^{*}-S$ relation. While we were able to 
derive this result in the low immigration regime, there was a small systematic 
derivation from the mean-field result and the simulation curves. 

While the stationary distribution is of considerable interest, the strength of
the approach that we have adopted here is that predictions of the time
evolution of the system are also possible. An approximation based on the number
of individuals in the system being large led to the picture of $P(n,t)$ as an
approximately Gaussian distribution broadening and moving with time. This
behavior may persist for quite a long time, especially when the immigration
rate is high, but eventually the Gaussian form is lost at large times. To
explore the approach to the stationary distribution a complimentary formalism
was required. Such a method was discussed in section V where a formal general
solution for the temporal evolution of the probability of having any species
represented by $n$ individuals is given. This solution $P(n,t)$ is given as a 
series expansion around the stationary state. In particular, such a solution 
allows one to predict quite well how the number of species in the system 
increases with time during the stochastic assembly process.
 
\bigskip

In summary, we believe that this simple model has illuminated the general
mechanisms at work in ecosystems and has allowed us to understand the broad
features of some of the universal phenomena seen in these systems. We hope that
the results presented here will motivate further work both in the increasingly 
sophisticated stochastic modelling of ecosystems and in the interpretation of
ecological data within a theoretical framework.

\acknowledgements
AJM wishes to thank the Complex Systems Research Group at the Universitat 
Polit\`ecnica de Catalunya, for hospitality while this work was been carried
out, and the British Council for support. This work has also been supported by 
a grant 1999FI 00524 UPC APMARN (DA) and by the Sante Fe Institute (RVS).

\appendix
\section{}
In this appendix we will give details of the derivations of the simpler
forms of the stationary distribution discussed in section III.

In our analysis, we will frequently make use of the asymptotic form for 
$\Gamma(z+a)/\Gamma(z)$ when $z \agt 1$ and $z \gg a$. Using Stirling's 
approximation for $z \agt 1$ one has that
\begin{displaymath}
\frac{\Gamma(z+a)}{\Gamma(z)} = z^{a} \left( 1 + 
\frac{a}{z} \right)^{(z+a - 1/2)}\, \exp{(-a)}
\end{displaymath}
\begin{displaymath}
\times \left[ 1 + P_{1}(z)\left(\frac{a}{z}\right) +  
P_{2}(z)\left(\frac{a}{z}\right)^{2} + ... \ \right]\, ,
\end{displaymath}
where the $P_{i}(z)$ are power series in $1/z$. Therefore, if in addition we
impose the condition $a \ll z$, then to a very good approximation
\begin{equation}
\frac{\Gamma(z+a)}{\Gamma(z)} \approx (1 + \epsilon )^{z(1+\epsilon)}\, z^{a} 
\, e^{-a}\ \ ; \ z \agt 1 \ ; \ \epsilon = \frac{a}{z} \ll 1\, .
\label{stirling}
\end{equation}
Note that $a$ need not be small, it simply has to be much less than $z$.

Applying (\ref{stirling}) to ${\cal F}_{3}(n,N)$ one finds 
($N \gg 1 \Rightarrow \nu^{*} \gg 1$)
\begin{eqnarray}
{\cal F}_{3}(n,N) & = & \left[ \frac{\Gamma(N + 1- n)}
{\Gamma(N + 1)} \right]^{-1} \left[ \frac{\Gamma(\nu^{*} - n)}
{\Gamma(\nu^{*})} \right] \nonumber \\ \nonumber \\
& \approx & \left( \frac{N + 1}{\nu^{*}}\right)^{n} \left\{ \left( 1 - 
\frac{n}{N + 1} \right)^{-(N + 1 - n)}\left( 1 -
\frac{n}{\nu^{*}} \right)^{(\nu^{*} - n)}\right\}\, . 
\label{approx_f3}
\end{eqnarray}
The term in curly brackets is equal to one, plus corrections which are 
negligible if $n_{\rm max}^{2}\lambda^{*}S/N^{2} \ll 1$ and 
$\lambda^{*}S \ll N$. Therefore, under these conditions
\begin{eqnarray}
{\cal F}_{3}(n,N) \approx \left( \frac{N}{\nu^{*}} \right)^{n} & \approx &
\exp{ -n \ln \left( 1 + \frac{\lambda^{*}S}{N} \right)} \nonumber \\ 
& \approx & \exp{ \left( -n\lambda^{*}S/N \right)}\, ,
\label{simple_f3}
\end{eqnarray}
using $S \gg 1$ and $\lambda^{*}S \ll N$ again.

To find a simpler form for ${\cal F}_{1}(n)$, we again apply (\ref{stirling}),
but with the more stringent condition $a \ll 1$. It then becomes 
$\Gamma(z+a)/\Gamma(z) \approx z^{a}$, and so for $\lambda^{*} \ll 1$ and
$n \agt 1$ 
\begin{equation}
{\cal F}_{1}(n) = 
\frac{1}{n\,\Gamma(\lambda^{*})}\, \frac{\Gamma(n + \lambda^{*})}{\Gamma(n)}
\approx \frac{\lambda^{*} n^{\lambda^{*}}}{n} = \frac{\lambda^{*}}{n}
\exp{\lambda^{*}\ln n}\,. 
\label{approx_f1}
\end{equation}
If we now ask that $\lambda^{*}\ln n_{\rm max} \ll 1$, we have that
\begin{equation}
{\cal F}_{1}(n) \approx \frac{\lambda^{*}}{n} \ \ ; \ n \agt 1 \ ; \ 
\lambda^{*} \ll \frac{1}{\ln n_{\rm max}}\, .
\label{simple_f1}
\end{equation}
To estimate $P_{s}(0)$ we apply (\ref{stirling}) directly to ${\cal F}_{2}(N)$ 
in (\ref{defeffs}). Assuming $\lambda^{*}S \agt 1$
\begin{eqnarray}
{\cal F}_{2}(N) & = & \left[ \frac{\Gamma(\nu^{*} + \lambda^{*})}
{\Gamma(\nu^{*})} \right]^{-1} \left[ \frac{\Gamma([\nu^{*} - N] + \lambda^{*})}
{\Gamma(\nu^{*} - N)} \right] \nonumber \\ \nonumber \\
& \approx & \left( \frac{\nu^{*} - N}{\nu^{*}}\right)^{\lambda^{*}} 
\left\{ \left( 1 + \frac{\lambda^{*}}
{\nu^{*}} \right)^{-(\nu^{*} + \lambda^{*})}\left( 1 +
\frac{\lambda^{*}}{\nu^{*} - N} \right)^{(\nu^{*} - N 
+ \lambda^{*})}\right\}\, . 
\label{approx_f2}
\end{eqnarray}
Under the very reasonable assumptions $(\lambda^{*})^{2} \ll N$ and 
$\lambda^{*} \ll S$, the curly brackets approximate very well to unity, and so
if in addition $\lambda^{*}S \ll N$,
\begin{equation}
{\cal F}_{2}(N) \approx \left( \frac{\lambda^{*}S}{N + \lambda^{*}S}
\right)^{\lambda^{*}} \approx \left( \frac{\lambda^{*}S}
{N}\right)^{\lambda^{*}} \approx \left( \mu^{*}S \right)^{\lambda^{*}}\, .
\label{simple_f2}
\end{equation}

This result may be used to find a useful expression for the average number of
species, $\langle S \rangle$ defined by (\ref{defavs}). Since 
$P_s(0) \approx \exp{\lambda^{*}\ln (\lambda^{*}S/N)}$, it follows that 
\begin{equation}
\langle S \rangle \approx \left( 1 - \exp{\lambda^{*}
\ln (\lambda^{*}S/N)} \right) S \approx S\lambda^{*}\ln 
\left( \frac{N}{S\lambda^{*}} \right)\, ,
\label{approx_S}
\end{equation}
if $ \lambda^{*}\ln (N/S\lambda^{*}) \ll 1$. We now write
\begin{eqnarray}
\ln \left( N/S\lambda^{*} \right) & = & \ln \left( [1 - \mu ] C^{*}/\mu \right)
\nonumber \\
& = & \epsilon^{-1} + \ln C^{*}\, , \ \ \ \ {\rm where\ } \epsilon^{-1} \equiv 
\ln \left( \frac{1 - \mu}{\mu} \right) \nonumber \\
& \approx & \epsilon^{-1} \exp {\epsilon \ln C^{*}}\, ,
\label{epsilon}
\end{eqnarray}
if $\epsilon|\ln C^{*}| \ll 1$. 

Now suppose that we assume that $\lambda^{*} \ll \epsilon$ and 
$\epsilon|\ln C^{*}| \ll 1$. It follows that 
$\lambda^{*}\epsilon^{-1}[1 + \epsilon|\ln C^{*}| ] \ll 1$ and therefore that
$\lambda^{*}\ln (N/S\lambda^{*}) \ll 1$. Thus this latter
condition may be replaced by $\lambda^{*} \ll \epsilon$ and 
$\epsilon|\ln C^{*}| \ll 1$. The immigration rate $\mu$ is typically much less
than one, so $\epsilon^{-1} \approx |\ln \mu |$. Therefore the last condition
becomes $| \ln C^{*} | \ll |\ln \mu |$. Putting all of this together, we find 
that
\begin{equation}
\langle S \rangle \approx \frac{\lambda^{*}S}{\epsilon}
\left( C^{*} \right)^{\epsilon} = \frac{N\mu}{(1 - \mu)\epsilon}
\left( C^{*} \right)^{-1 + \epsilon}\, ,
\label{simple_S}
\end{equation}
if $\lambda^{*}S \agt 1$ (since we have used (\ref{simple_f2})), 
$\lambda^{*} \ll \epsilon$ and $|\ln C^{*}| \ll |\ln \mu|$.

\section{}

In this appendix we give details of the calculations presented in section V.
We begin by showing that is we apply the conditions (\ref{conds_F}) to
the general solution (\ref{gensoln}) of the partial differential equation 
(\ref{pde}), we obtain (\ref{final_form}) with the constants $\{w_{k}\}$
being determined by (\ref{finding_wk}).

First, let us apply the condition $F(1,t) = 1$. Now
$\Phi^{(1)}_{\lambda} \approx (1 - z)^{N+a}$ and 
$\Phi^{(2)}_{\lambda} \approx (1 - z)^{N+b}$ as $z \rightarrow 1$. So define
$\tilde{a} = N + a$ and $\tilde{b} = N + b$. Then from (\ref{abc}),
\begin{eqnarray}
\tilde{a} & = & \frac{\left[ 1 - \lambda^{*}S \right] - \sqrt{\left[ 1
- \lambda^{*}S \right]^{2} + 4\lambda}}{2}\nonumber \\
\tilde{b} & = & \frac{\left[ 1 - \lambda^{*}S \right] + \sqrt{\left[ 1
- \lambda^{*}S \right]^{2} + 4\lambda}}{2}\, ,
\label{tildes}
\end{eqnarray}
with $\tilde{a}\tilde{b} = - \lambda$. Since the constants $a$ and $b$ appear
in the differential equation (\ref{hyper}) symmetrically, we have made a 
choice as to which has the positive and which has the negative square root.
From the general theory of the master equation~\cite{van}, the eigenvalues 
$\lambda$ are real and non-negative, and so $\tilde{a}$ and $\tilde{b}$ are 
real. Moreover, if $\lambda \neq 0$, they are of different signs, since their 
product is negative. With the choice (\ref{tildes}), $\tilde{a} < 0$ and 
$\tilde{b} > 0$. From these results we deduce that 
$\Phi^{(1)}_{\lambda}$ diverges and $\Phi^{(2)}_{\lambda} \rightarrow 0 $ as 
$z \rightarrow 1$. We must therefore take $v_{\lambda} = 0$ for all 
$\lambda > 0$. When $\lambda = 0$, $\tilde{b} = 1 - \lambda^{*}S$, which may 
be negative, and so we also take $w_{0} = 0$. Finally, $\tilde{a} = 0$ when 
$\lambda = 0$ and so this term is the only one which is not zero or does not 
diverge as $z \rightarrow 1$. Since the $\lambda = 0$ solution is the 
stationary solution, the condition $F(1,t) = 1$ is automatically satisfied as 
long as the stationary solution is normalized. Therefore, the application of 
this condition has reduced (\ref{gensoln}) to
\begin{equation}
F(z,t) = F_{s}(z) + \sum_{\lambda > 0}\, w_{\lambda}\, (1 - z)^{N+b}\,
F(b,b - c + 1, b - a + 1; 1 - z)\, e^{-\lambda \zeta t}\, ,
\label{lessgensoln}
\end{equation}
where $F_{s}(z) = \sum_{n}P_{s}(n)z^{n}$.

Before proceeding any further, we need to investigate the sum over $\lambda$ 
more carefully. We know that this sum should be over a set of discrete
integers: $F(z,t) = \sum_{n}P(n,t)z^{n}\, , \ n \in \{0,1,\ldots,N\}$. An
analysis of the structure of the hypergeometric function in (\ref{lessgensoln})
for large $z$ shows that this will only be so if $b$ is equal to an integer
which is zero or negative: $b = - l$. We can understand this condition by 
recalling~\cite{abr} that the function $F(a',b',c';x)$ is a polynomial of
degree $l$ (where $l$ is a non-negative integer) in $x$ if $a' = - l$.
Therefore, if $b = -l$, then $F$ in (\ref{lessgensoln}) must be a polynomial
of degree $l$ in $(1 - z)$, i.e. $(1 - z)^{N+b}F$ must be a polynomial of
degree $N+b+l = N$ in $(1 - z)$, as required.

From (\ref{tildes}), $\tilde{a} + \tilde{b} = \left[ 1 - \lambda^{*}S\right]$,
so if $\tilde{b} = N + b = (N - l)$, then 
$\tilde{a} = l -N + \left[ 1 - \lambda^{*}S\right]$. Similarly, 
$\tilde{b} - \tilde{a} = 2N - 2l - \left[ 1 - \lambda^{*}S \right] =
\sqrt{[1 - \lambda^{*}S]^{2} + 4\lambda}$. A short calculation then gives
$\lambda = (N - l)(\omega - l)$, where $\omega = N + \lambda^{*}S - 1$. Since
we require $\lambda > 0$ in the sum in (\ref{lessgensoln}), then 
$l = 0,1,\ldots,N-1$. So, in summary,
\begin{eqnarray}
\lambda = (N - l)(\omega - l)\ \ & ; & \ \ l=0,1,\ldots,N-1 \nonumber \\
a = l - N - \omega \ \ ; \ \ b & = & - l\ \ ; \ \ c = 1 - \lambda^{*} - N\, ,
\label{lambda}
\end{eqnarray}
where
\begin{equation}
\omega \equiv N + \lambda^{*}S - 1\, .
\label{omega}
\end{equation}
Rather than $l$, it is preferable to use $k \equiv N - l$ to label the 
time-dependent solutions. Then (\ref{lessgensoln}) becomes
\begin{equation}
F(z,t) = F_{s}(z) + \sum_{k=1}^{N}\, w_{k}\, (1 - z)^{k}\,
F(k - N,k + \lambda^{*},2k + \lambda^{*}S; 1 - z)\, 
e^{-k(k + \lambda^{*}S - 1) \zeta t}\, ,
\label{discreteform1}
\end{equation}
where we have written $w_{\lambda(l)}$ as $w_{k}$ for convenience. We note that
if there were a $k=0$ term in the sum, it would equal 
$w_{0}\,F(- N,\lambda^{*},\lambda^{*}S; 1 - z)$ which in turn equals~\cite{abr}
\begin{displaymath}
w_{0}\,\frac{\Gamma(\nu^{*}) \Gamma(\nu^{*} + \lambda^{*} - N)}
{\Gamma(\nu^{*} - N) \Gamma(\nu^{*} + \lambda^{*})}\, 
F(-N,\lambda^{*},1 - \nu^{*}; z)
\end{displaymath}
\begin{equation}
= w_{0}\,\sum_{n=0}^{N}\, {N \choose n} {\Gamma (n + \lambda^*) \over \Gamma(
\lambda^*)} {\Gamma (\nu^* - n) \over \Gamma(\nu^* - N)} 
{\Gamma (\lambda^* + \nu^* - N) \over \Gamma(\lambda^* + \nu^*)}\, z^{n}\, 
\label{kequalszero}
\end{equation}
which equals $w_{0}\,\sum_{n}P_{s}(n)z^{n}$, using (\ref{Ps(n)_gamma}).
Therefore (\ref{discreteform1}) may be written in the alternative form
\begin{equation}
F(z,t) = \sum_{k=0}^{N}\, w_{k}\, (1 - z)^{k}\,
F(k - N,k + \lambda^{*},2k + \lambda^{*}S; 1 - z)\, 
e^{-k(k + \lambda^{*}S - 1) \zeta t}\, ,
\label{discreteform2}
\end{equation}
where $w_{0} = 1$. It is also possible to write the function $F$ in 
(\ref{discreteform2}) as a Jacobi polynomial of order $N - k$~\cite{abr}.

The final step is to apply the initial condition given in (\ref{conds_F}) to
(\ref{discreteform2}) in order to determine the sets of constants $\{w_{k}\}$.
This leads to 
\begin{equation}
z^{m} = \sum_{k=0}^{N}\, w_{k}\, (1 - z)^{k}\,
F(k - N,k + \lambda^{*},2k + \lambda^{*}S; 1 - z)\, .
\label{initialcond}
\end{equation}
To solve (\ref{initialcond}), let us write the hypergeometric function as
\begin{equation}
F = \sum_{n=0}^{N - k}\, f_{n}(1- z)^{n}\ \ \ {\rm where}\ \ 
f_{n} = (-1)^{n}\, {N - k \choose n}\, \frac{(k + \lambda^{*})_{n}}
{(2k + \lambda^{*}S)_{n}}\, ,
\label{terminating}
\end{equation}
where the symbol $(a)_n$ means $\Gamma(a + n)/\Gamma(a)$. Then rearranging the 
double sum gives
\begin{equation}
z^{m} = \sum^{N}_{n=0}\, (1 - z)^{n} \sum^{n}_{k=0}\, w_{k}\, f_{n-k}
= \sum^{N}_{n=0}\, c_{n}\, (1 - z)^{n}\, ,
\label{rearrange}
\end{equation}
where $c_{n} = \sum^{n}_{k=0}\, w_{k}\, f_{n-k}$. However, it is
straightforward to determine the $c_{n}$: by expanding $(1 - [1 - z])^{m}$ in
powers of $(1 - z)$ we find that $c_n$ vanishes for $n > m$ and is 
proportional to a binomial coefficient for $n \le m$. Using this result and 
writing out $f_{n-k}$ fully yields
\begin{equation}
\sum^{n}_{k=0}\, w_{k}\, (-1)^{n-k}\, {N-k \choose n-k}\, 
\frac{\Gamma(n + \lambda^{*})\, \Gamma(2k + \lambda^{*}S)}
{\Gamma(k + \lambda^{*})\, \Gamma(n + k + \lambda^{*}S)} =
\left\{ \begin{array}{ll} 
(-1)^{n}\,{m \choose n}, & \mbox{\ if $n \le m$} \\ 
\ \ \ \ \ \ 0 \ \ \ \ \ \ , & \mbox{\ if $n > m$\, .}
\end{array} \right.
\label{implicit}
\end{equation}

We now turn to the problem of finding $P(n,t)$, which involves identifying the
the coefficient of $z^{n}$ in (\ref{discreteform2}). Let us begin by 
considering $F(k - N,k + \lambda^{*},2k + \lambda^{*}S; 1 - z)$. By following 
exactly the same steps that lead to (\ref{kequalszero}) in the $k=0$ case, but 
this time for general $k$, we find that this function equals
\begin{displaymath}
\frac{\Gamma(\nu^{*}) \Gamma(2k + \nu^{*} + \lambda^{*} - N)}
{\Gamma(k + \nu^{*} - N) \Gamma(k + \nu^{*} + \lambda^{*})}\, 
F(k - N,k + \lambda^{*},1 - \nu^{*}; z)
\end{displaymath}
\begin{equation}
= \sum_{r=0}^{N-k}\, {N-k \choose r} {\Gamma (r + k + \lambda^*) \over \Gamma(
k + \lambda^*)} {\Gamma (\nu^* - r) \over \Gamma(k + \nu^* - N)} 
{\Gamma (2k + \lambda^* + \nu^* - N) \over \Gamma(k + \lambda^* + \nu^*)}\, 
z^{r}\, , 
\label{generalk}
\end{equation}
Therefore, we may write the solution for $F(z,t)$, eqn. (\ref{discreteform2}),
as
\begin{equation}
F(z,t) = \sum^{N}_{k=0}\, w_{k} (1 - z)^{k}\, \sum^{N-k}_{r=0}\, \rho_{r}(k)
z^{r} e^{-k(k + \lambda^{*}S -1)\zeta t}\, ,
\label{double_sum}
\end{equation}
where
\begin{equation}
\rho_{r}(k) = {N-k \choose r} {\Gamma (r + k + \lambda^*) \over \Gamma(
k + \lambda^*)} {\Gamma (\nu^* - r) \over \Gamma(k + \nu^* - N)} 
{\Gamma (2k + \lambda^* + \nu^* - N) \over \Gamma(k + \lambda^* + \nu^*)}\, .
\label{def_rho}
\end{equation}
To identify powers of $z$ in (\ref{double_sum}) we break it down further:
\begin{equation}
F(z,t) = \sum^{N}_{k=0}\, w_{k}\,e^{-k(k + \lambda^{*}S -1)\zeta t}\, 
\sum^{k}_{l=0}\, (-1)^{l} {k \choose l}\, \sum^{N-k}_{r=0}\, \rho_{r}(k)
z^{r+l}\, .
\label{triple_sum}
\end{equation}
Writing
\begin{equation}
\sum^{k}_{l=0}\, (-1)^{l} {k \choose l}\, \sum^{N-k}_{r=0}\, \rho_{r}(k)
z^{r+l} = \sum^{N}_{n=0}\, \chi(n,k)\, z^{n}\, ,
\label{def_chi}
\end{equation}
we have from (\ref{triple_sum})
\begin{eqnarray}
F(z,t) & = & \sum^{N}_{k=0}\, w_{k}\,e^{-k(k + \lambda^{*}S -1)\zeta t}\, 
\sum^{N}_{n=0}\, \chi(n,k)\, z^{n} \nonumber \\ \nonumber \\
& = & \sum^{N}_{n=0}\, \left\{ \sum^{N}_{k=0}\, w_{k}\,\chi(n,k)\,
e^{-k(k + \lambda^{*}S -1)\zeta t} \right\} z^{n} \nonumber \\ \nonumber \\
\Rightarrow \ \ \ P(n,t) & = & \sum^{N}_{k=0}\, w_{k}\,\chi(n,k)\,
e^{-k(k + \lambda^{*}S -1)\zeta t}\, .
\label{FtoP}
\end{eqnarray}
So to find $P(n,t)$ we have to determine $\chi(n,k)$, which is the coefficient
of $z^{n}$ in (\ref{def_chi}). If we denote $(-1)^{l}{k \choose l}$ by $a_{l}$
and $\rho_{r}(k)$ by $b_{r}$, this is like asking: what is the coefficient of 
$z^{n}$ in 
\begin{displaymath}
\left[ a_{0} + a_{1}z + \ldots + a_{k}z^{k} \right]
\left[ b_{0} + b_{1}z + \ldots + b_{N-k}z^{N-k} \right]\, ?
\end{displaymath}
The answer is:
\begin{displaymath}
\sum^{r={\rm min}\{N-k,n\}}_{r={\rm max}\{n-k,0\}}\, a_{n-r}b_{r}\, .
\end{displaymath}
Going back to the variables relevant to the problem under consideration,
\begin{equation}
\chi(n,k) = \sum^{r={\rm min}\{N-k,n\}}_{r={\rm max}\{n-k,0\}}\, 
(-1)^{n-r}\, {k \choose n-r}\, \rho_{r}(k)\, .
\label{chi}
\end{equation}
The complicated limits on the sum can be relaxed with a suitable interpretation
of the binomial coefficient ${M \choose L}$. For instance,
\begin{displaymath}
(-1)^{n-r}\,{k \choose n-r} = \frac{\Gamma(-k + [n-r])}{(n-r)\,!\, 
\Gamma(-k)}\, .
\end{displaymath}
If $(n-r) \le k$, both the numerator and the denominator on the right-hand-side
diverge in such a way that the ratio is finite and non-zero and is commonly
written as the left-hand-side. If $(n-r) > k$, then only the denominator
diverges and the right-hand-side is zero. With this understanding, the lower 
condition on the sum in (\ref{chi}) can simply be replaced by $r=0$, since
there is no contribution if $(n-r) > k$, that is, if $r < n-k$. A similar 
interpretation of the binomial coefficient in $\rho_{r}(k)$ can be used to 
remove the upper condition on the sum. The result (\ref{chi}), together with 
the definition of $\rho_{r}(k)$, determine the $P(n,t)$ given by (\ref{FtoP}).

\end{document}